\documentclass[aps,prx,reprint,longbibliography]{revtex4-1}

\usepackage{color}
\usepackage{graphicx}
\usepackage{amsmath}
\usepackage{bm}
\usepackage{amsfonts}
\usepackage[mathscr]{euscript}

\newcommand{\bket}[1]{\left<#1\right>}

\newcommand{\ket}[1]{\left|#1\right>}
\newcommand{\bra}[1]{\left<#1\right|}

\newcommand{\g}{\ket{g}}
\newcommand{\e}{\ket{e}}
\newcommand{\f}{\ket{f}}

\begin{document}
\raggedbottom

\title{Robust concurrent remote entanglement between two superconducting qubits}

\author{A. Narla, S. Shankar, M. Hatridge, Z. Leghtas, K. M. Sliwa, E. Zalys-Geller, S. O. Mundhada, W. Pfaff, L. Frunzio, R. J. Schoelkopf, M. H. Devoret}

\affiliation{Department of Applied Physics, Yale University}

\date{\today}

\begin{abstract}

Entangling two remote quantum systems which never interact directly is an essential primitive in quantum information science and forms the basis for the modular architecture of quantum computing. When protocols to generate these remote entangled pairs rely on using traveling single photon states as carriers of quantum information, they can be made robust to photon losses, unlike schemes that rely on continuous variable states.  However, efficiently detecting single photons is challenging in the domain of superconducting quantum circuits because of the low energy of microwave quanta. Here, we report the realization of a robust form of concurrent remote entanglement based on a novel microwave photon detector implemented in the superconducting circuit quantum electrodynamics (cQED) platform of quantum information. Remote entangled pairs with a fidelity of $0.57\pm0.01$ are generated at $200$~Hz. Our experiment opens the way for the implementation of the modular architecture of quantum computation with superconducting qubits.

\end{abstract}


\maketitle

\clearpage

\section{Introduction}

The concept of a photon, the quantum of excitation of the electromagnetic field, was introduced by Planck and Einstein to explain the black-body radiation spectrum\cite{Planck1901} and the photoelectric effect\cite{Einstein1905}. However, experiments that would definitively prove the existence of traveling optical photons as independent entities were only understood\cite{Glauber1963, Cohen-Tannoudji1998} and realized\cite{Hong1987} much later in the $20^{\textrm{th}}$ century. Although there is no reason to suppose that microwave photons would behave differently than their optical counterparts, revealing and manipulating them is challenging because their energies are $4$ to $5$ orders of magnitude lower. Cavity-QED, and later on circuit-QED, have established the reality of stationary quantum microwave excitations of a superconducting resonator by strongly coupling them to Rydberg\cite{Haroche2006} and superconducting artificial atoms\cite{Schuster2007}. The production of traveling microwave photons was then indirectly demonstrated using linear amplifiers to measure the state of the radiation\cite{Houck2007, Bozyigit2011, Lang2013}. However, while there have been proposals and implementations of single flying microwave photon detectors\cite{Chen2011, Fan2014, Inomata2016}, controlling and employing the single-photon nature of microwave radiation is still an open challenge. Here, we carry over to the microwave domain the remote entanglement experiment performed in quantum optics by realizing and operating a single photon detector based on a superconducting 3D transmon qubit\cite{Paik2011}.

With single microwave photon detectors still not commonly used, the only form of remote entanglement realized so far with superconducting qubits has been through the use of continuous variable coherent states as the flying information carriers\cite{Roch2014}. While such states can be efficiently synthesized by standard microwave equipment and processed by quantum-limited linear parametric amplifiers\cite{Castellanos-Beltran2007, Bergeal2010a} readily available at microwave frequencies, the disadvantage is this route is its sensitivity to losses in the paths of the flying states. In contrast, remote entanglement using flying single photons is robust to these losses, as demonstrated in the optical domain\cite{Moehring2007, Hofmann2012, Bernien2013, Hensen2015, Delteil2015}. This protocol offers the advantage that only the successful detection of photons is linked to the production of a pure entangled state\cite{Cabrillo1999, Barrett2005}. This feature is particularly important for generating entanglement between two distant stationary qubits, a crucial element of the modular architecture of quantum computation\cite{Monroe2014} and the proposed quantum internet\cite{Kimble2008}. Furthermore, scaling up the modular architecture requires no direct connections between modules, unlike previously demonstrated sequential methods\cite{Roch2014}, maintain a strong on/off ratio. Thus, demonstrating robust remote entanglement which satisfies this requirement, i.e. a \textit{concurrent} protocol, is a vital step in the implementation of the modular architecture with superconducting qubits.

\begin{figure*}[!ht]
\includegraphics{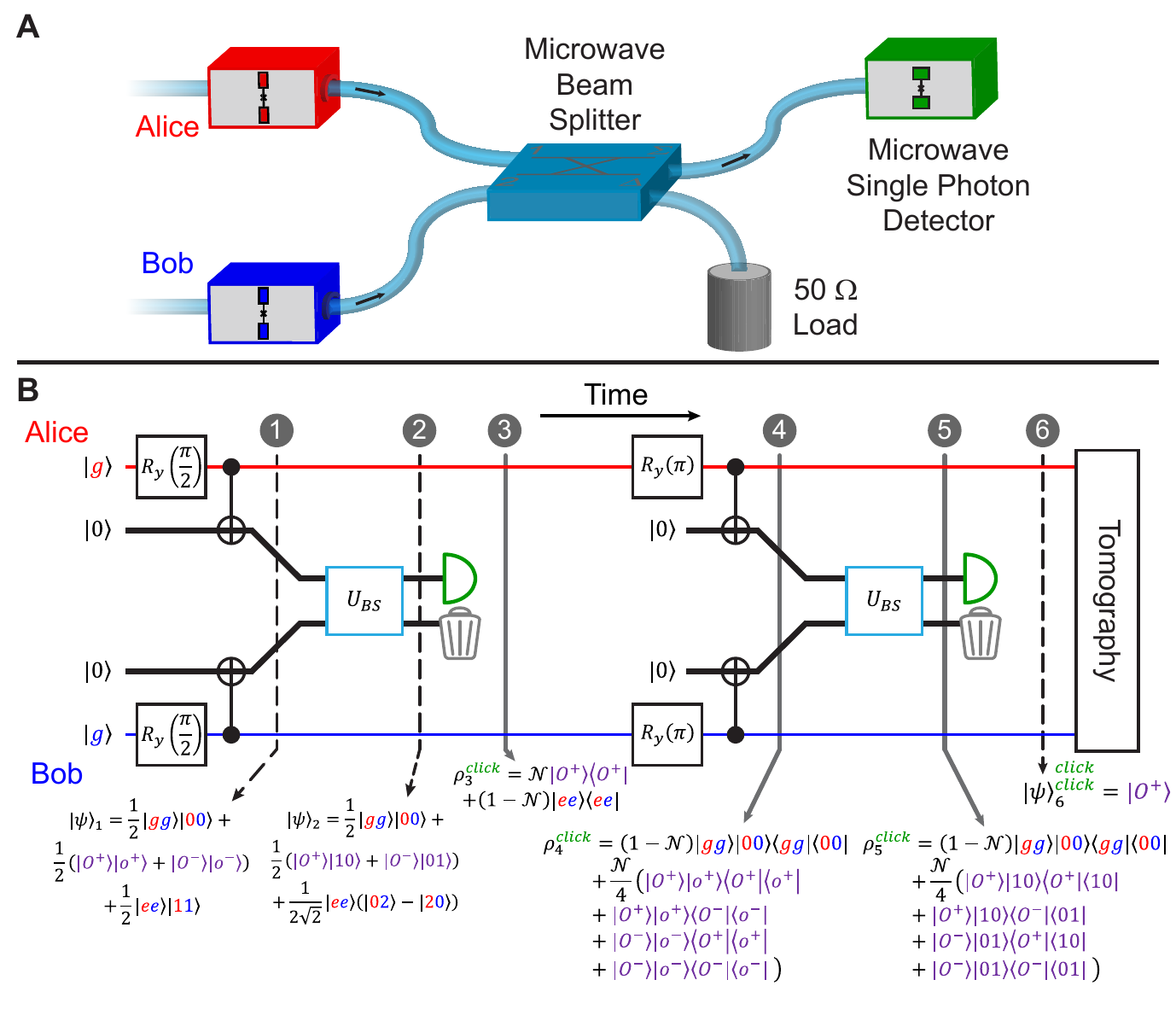}
\caption{\textbf{$\mid$ Experiment and protocol schematic for remote entanglement of transmon qubits using flying single microwave photons.} 
A) Two superconducting 3D transmon qubits, Alice and Bob, are connected by coaxial cables to the two input ports of the microwave equivalent of a $50/50$ beam-splitter. One of the output ports of the splitter is connected to a microwave single photon detector also realized by a 3D transmon qubit. The other port of the splitter is terminated in a cold $50~\Omega$ load. 
B) Quantum circuit diagram of the remote entanglement protocol, with the states of the quantum system at various steps. The Alice and Bob (red and blue) qubits are each prepared in the state $\frac{1}{\sqrt{2}}\left(\g+\e\right)$ by a single qubit gate $R_y\left(\frac{\pi}{2}\right)$. They are then entangled with flying single photons (black) via a CNOT-like operation. The states $\ket{O^{\pm}}=\frac{1}{\sqrt{2}}\left(\ket{ge}\pm\ket{eg}\right)$ represent odd Bell states of the Alice and Bob qubits while $\ket{o^{\pm}}=\frac{1}{\sqrt{2}}\left(\ket{10}\pm\ket{01}\right)$ represent odd Bell states of flying single photons in the Alice and Bob channels respectively. 
The flying photons interfere on the beam-splitter whose unitary action $\bm{U_{\textrm{BS}}}$ erases their which-path information. Following a $\pi$-pulse on Alice and Bob, the CNOT-like operation and beam-splitter steps are repeated to remove contributions of the unwanted $\ket{ee}$ state. Detecting two photon clicks in a pair of consecutive rounds heralds the $\ket{O^+}=\frac{1}{\sqrt{2}}\left(\ket{ge}+\ket{eg}\right)$ Bell state of Alice and Bob.}
\end{figure*}

\section{Overview of Experiment and Protocol}

The experiment, housed in a dilution refrigerator below $20$~mK, consists of two different superconducting transmon qubits (see Fig.~1A), referred to as Alice and Bob, in separate 3D cavities\cite{Paik2011}. The cavities have nearly identical resonance frequencies $\omega_{A}^g/2\pi = 7.6314~\textrm{GHz}$, $\omega_{B}^g/2\pi = 7.6316~\textrm{GHz}$  and bandwidths $\kappa_A/2\pi = 1.2~\textrm{MHz}$, $\kappa_B/2\pi = 0.9~\textrm{MHz}$. Their strongly coupled output ports are connected by microwave coaxial cables to the two input ports of a $180^\circ$ hybrid, the microwave equivalent of a $50/50$ beamsplitter. One of the output ports of the hybrid is connected to a microwave single photon detector which is realized by a third 3D cavity also containing a transmon. The other output port of the hybrid is terminated in a $50~\Omega$ load. To ensure signal flow as shown by the arrows in Fig.~1A, microwave isolators/circulators (not shown, see experimental schematic in Appendix A) are inserted into the lines connecting each qubit to the hybrid. These provide robust isolation between modules and connect the system output to readout electronics.

To entangle the remote qubits, flying microwave single photon states are used as carriers of quantum information according to the protocol proposed in \cite{Barrett2005}. As outlined in Fig.~1B, the remote entanglement protocol begins by initializing both qubit-cavity systems in $\frac{1}{\sqrt{2}}\left(\g+\e\right)\otimes\ket{0}$, the state on the equator of the Bloch sphere with no photons in their respective cavities. Through a controlled-NOT (CNOT)-like operation, whose implementation is detailed later in the text, the qubits are now entangled with flying single photons where the state of each qubit-photon pair becomes $\frac{1}{\sqrt{2}}\left(\ket{g0}+\ket{e1}\right)$. The joint state of all stationary and flying qubits can be expressed as $\ket{\psi}_1=\frac{1}{2}\left(\ket{gg}\ket{00} + \ket{O^+}\ket{o^+} + \ket{O^-}\ket{o^-} + \ket{ee}\ket{11}\right)$ where $\ket{O^{\pm}}=\frac{1}{\sqrt{2}}\left(\ket{ge}\pm\ket{eg}\right)$ represent the odd Bell states of the Alice and Bob qubits and $\ket{o^{\pm}}=\frac{1}{\sqrt{2}}\left(\ket{10}\pm\ket{01}\right)$ represent the odd Bell states of flying single photons in Alice's and Bob's channels, respectively. The photons interfere on the $180^\circ$ hybrid whose action, analogous to that of a beam-splitter, is described by the unitary $\bm{U_{\textrm{BS}}} = e^{-3\pi\left(\bm{a}^{\dagger}\bm{b} - \bm{a}\bm{b}^{\dagger}\right)/4}$. This maps $\ket{o^+} \rightarrow \ket{10}$ ($\ket{o^-} \rightarrow \ket{01}$), taking the two flying odd Bell states to a single photon state in the Alice or Bob branch of the detector part of the system. This operation erases the which-path information of the photons and produces Hong-Ou-Mandel interference\cite{Hong1987}. After the hybrid, the total system state is $\ket{\psi}_2=\frac{1}{2}\left(\ket{gg}\ket{00} + \ket{O^+}\ket{10} + \ket{O^-}\ket{01} + \frac{1}{\sqrt{2}}\ket{ee}\left(\ket{02} - \ket{20}\right)\right)$. At this point, the photons in the Alice channel enter the detector which distinguishes between detecting a photon, a `click', or detecting nothing, called `no click'. Ideally, by heralding on only single photon detection events, the $\ket{O^+}$ is selected out from all the other states. However, losses in the system between the qubits and the detector and the inability of the detector to distinguish between the Fock states $\ket{1}$ and $\ket{2}$ instead result in the mixed density matrix $\rho_3^{click}=\mathscr{N} \ket{O^+}\bra{O^+} + \left(1-\mathscr{N}\right) \ket{ee}\bra{ee}$ when the detector clicks. Here, the normalization constant $\mathscr{N}$ depends on loss in the system and the characteristics of the detector (see Appendix F). In particular, it depends on the probabilities with which it maps the input flying photon states, $\ket{1}$ and $\ket{2}$, to an outcome of click. Another crucial assumption in $\rho_3^{click}$ is that the detector has no dark counts, i.e. it never clicks when it receives $\ket{0}$. A fuller version of $\rho_3^{click}$ including dark counts is given in the Appendix F. Thus, at this stage, the qubits are in the state $\ket{O^+}$ with probability $\mathscr{N}$ and we would like to remove the undesired $\ket{ee}$ state.

To achieve this, a $R_y(\pi)$ pulse is applied on both Alice and Bob followed by a second round of entangling the qubits with flying photons, interfering them on the hybrid and detecting them. The $\pi$-pulse takes $\ket{ee} \rightarrow \ket{gg}$; consequently, in the second round, the unwanted state is mapped onto $\ket{gg}\ket{00}$, and thus it can be selected out by detecting a photon. On the other hand, $\ket{O^+}$ is mapped onto a superposition of $\ket{O^+}\ket{10}$ and $\ket{O^-}\ket{01}$. Conditioning on measuring clicks in two consecutive rounds of the protocol results in the odd Bell state $\ket{\psi}_6^{click,~click}=\ket{O^+}$. A result of this dual conditioning is that losses in the system only reduces the success probability of the protocol and not the fidelity of the generated entangled state. Replacing the cold $50~\Omega$ load with a second detector would increase the success probability by a factor of $4$ and allows for the generation of both the $\ket{O^+}$ and $\ket{O^-}$ states depending on whether the same or different detectors go click on each round, respectively. Since it does not improve the fidelity of entanglement, we omitted the second detector to simplify the microwave control electronics and cold hardware.

\begin{figure*}[!ht]
\includegraphics{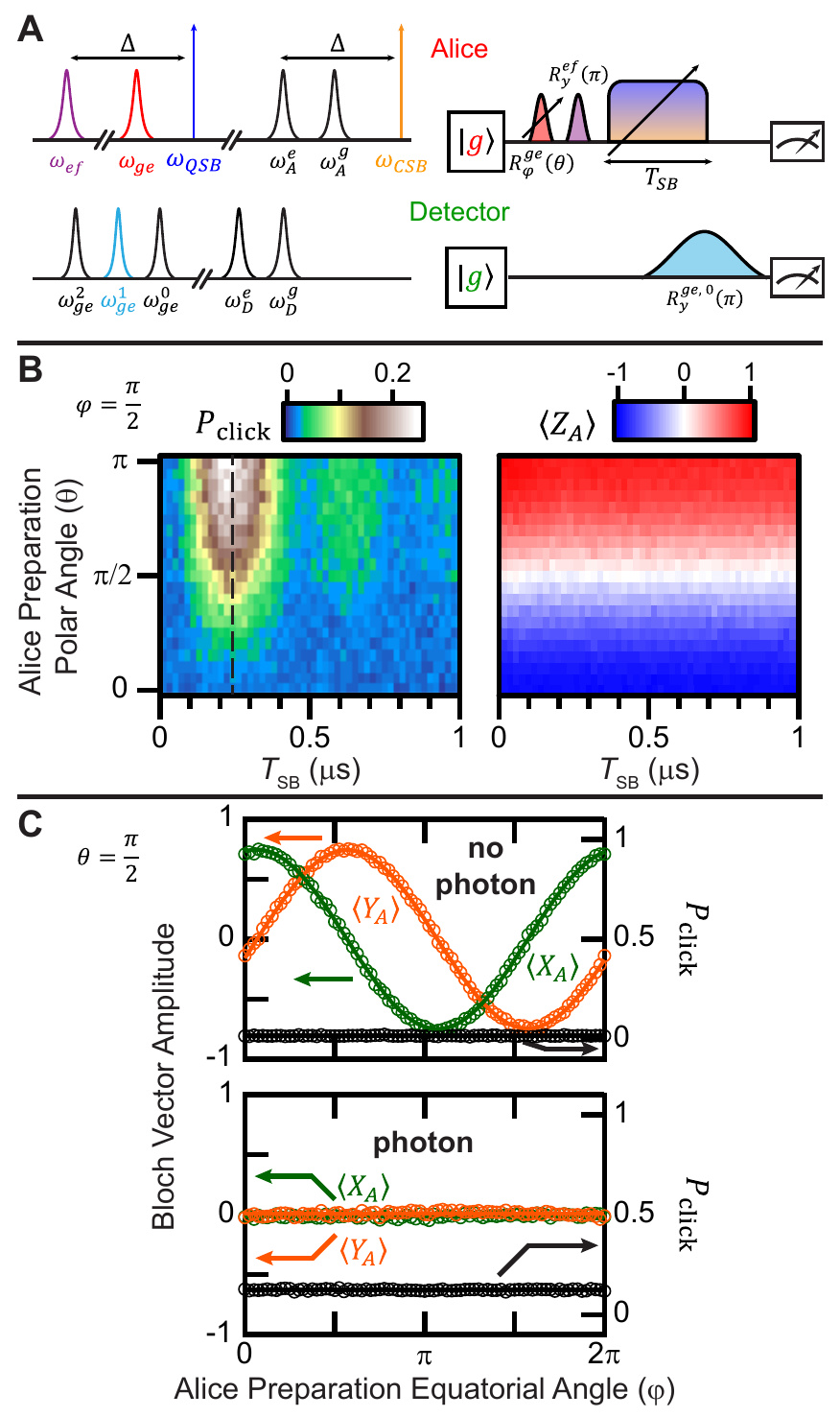}
\caption{\textbf{$\mid$ Signatures of qubit/flying photon entanglement.}
A) Frequency spectra of the Alice and detector qubit-cavity systems (left) and experimental pulse sequence (right). The colors denote transitions which are driven to perform the CNOT-like operation and flying single photon detection. The Alice qubit is prepared in an arbitrary initial state by the pulse $R_{\phi}^{ge}\left(\theta\right)$ at $\omega_{ge}$. The CNOT-like operation consists of a $R_y\left(\pi\right)$ pulse at $\omega_{ef}$ followed by a pair of sideband pulses. The sideband pulses are applied at  $\omega_{QSB}$, detuned by $\Delta$ from $\omega_{ef}$, and $\omega_{CSB}$, detuned by $\Delta$ from $\omega_{A}^e$. To detect flying photons, a frequency selective $\pi$-pulse is applied to the detector qubit at $\omega_{ge}^1$ followed by a measurement of the qubit state. 
B) Color plots of the probability, $P_{\textrm{click}}$, of the detector qubit ending in $\e$(left) and the Alice qubit polarization, $\bket{Z_A}$ (right), as a function of the sideband pulse length $T_{\textrm{SB}}$ and $\theta$ (for $\phi=\pi/2$). The dashed line at $T_{\textrm{SB}}=254$~ns corresponds to a transfer $\ket{f0}\rightarrow\ket{e1}$, i.e. a CNOT-like operation.
C) Detector click probability, $P_{\textrm{click}}$, and Alice equatorial Bloch vector components, $\bket{X_A}$ and $\bket{Y_A}$, as a function of $\phi$ for $\theta=\pi/2$ when the CNOT-like operation is either performed (bottom) or not (top). Open circles are experimental data and lines are fits.}
\end{figure*}

Successfully realizing this protocol required simultaneously: (1) implementing the generation of single photon Fock states which are entangled with the stationary qubits and (2) detecting the subsequent single photon states. Furthermore, the frequencies and temporal envelopes of the photons arising from each cavity had to be controlled to ensure that the detector cannot distinguish between them. 

\section{Description of the Qubit-Photon Entanglement Process}

The first ingredient, previously termed a CNOT-like operation, actually maps an arbitrary qubit state $\alpha\ket{g0} + \beta\ket{e0}$, where $\alpha$ and $\beta$ are arbitrary complex coefficients, onto the joint qubit-flying photon state $\alpha\ket{g0} + \beta\ket{e1}$ (this operation is not a unitary in the manifold $\left\{\ket{g0}, \ket{g1}, \ket{e0}, \ket{e1}\right\}$ because it takes $\ket{e1}$ to $\ket{f1}$; however, this has no effects on the protocol since the cavity always starts in $\ket{0}$). This is done by exploiting $\ket{f}$, the second excited state of the transmon qubit\cite{Koch2007}, as well as the two-photon transition $\ket{f0} \leftrightarrow \ket{e1}$\cite{Kindel2015, Pechal2014}. As shown in Fig.~2A, starting with the qubit in $\alpha\g + \beta\e$, the operation is realized by first applying a $\pi$-pulse at $\omega_{ef}^{0}$, taking the qubit to $\alpha\g + \beta\f$, and then applying a $\pi$-pulse on the $\ket{f0} \leftrightarrow \ket{e1}$ with two sideband tones ($\omega_{QSB}$, $\omega_{CSB}$). This maps the qubit state onto the joint qubit-intra-cavity state, $\alpha\ket{g0} + \beta\ket{e1}$. Finally, the photon state leaks out of the cavity, becoming a flying state that is entangled with the qubit. As a result, the traveling photon has the frequency $\omega_{A}^e$ ($\omega_{B}^e$) and a decaying exponential temporal waveform with the decay constant $\kappa_{A}$ ($\kappa_{B}$). The indistinguishability of the photons, then, was achieved in this experiment by the nearly identical frequencies and bandwidths of the Alice and Bob cavities (as given above and further discussed in Appendix C). Note that although the photons need to overlap in frequency, there is no requirement here for the qubits to be identical.

\section{Description of Microwave Single Photon Detection}

The second ingredient of the experiment, microwave single photon detection, is the novel technical component of our demonstration. Put simply, this detector is just another transmon-3D cavity system like Alice and Bob. The strongly coupled port of the cavity is the detector input port. In the strong dispersive regime where the qubit is operated ($\chi_D/2\pi=3$~MHz, $\kappa_D/2\pi=1$~MHz), we can selectively $\pi$-pulse the qubit conditioned on the presence of one intra-cavity photon\cite{Schuster2007}, mapping the flying photon onto the state of the detector qubit. To operate this system as a detector of single flying photons, we tuned the cavity frequency $\omega_{D}^{g}/2\pi=7.6222$~GHz close to $\omega_{A}^e$ and $\omega_{B}^e$ and matched the linewidths of all three cavities. This condition ensured that the detector efficiency is maximized. The incident single photons from Alice and Bob will excite the detector cavity $\sim50\%$ of the time (see Appendix D) since their decaying exponential temporal waveforms are not mode-matched to the cavity. Thus, the selective $\pi$-pulse excites the qubit only if a photon was received, with the length and timing of this pulse determining the detector efficiency (see Appendix D). Once the photon leaks back out, a conventional cQED dispersive readout of the qubit state\cite{Blais2004} completes the quantum non-demolition (QND) photon detection process. Measuring the qubit in the excited state corresponds to a photon detection event (click). Finally, the detector is reset by returning the qubit to $\g$ with an un-selective $\pi$-pulse.

This microwave photon detector satisfies three important criteria in an architecture that is easily integrated with current state-of-the art cQED experiments. First, the detector has a reasonable efficiency, $\eta\approx0.5$, since about half of all incident photons enter the detector. Second, this detector has low dark counts (the probability of the detector reporting a click even when no photon entered the detector) $P_{\textrm{d}}<0.01$, limited by the frequency selectivity of the $\pi$-pulse. Finally, it has a short re-arm time of $450$~ns determined by how long it takes to empty the cavity and reset the qubit. We discuss avenues to further improving this detector in Appendix D of the paper. Nevertheless, as we show below, the detector performance is sufficient for generating remote entanglement.

\begin{figure*}[!ht]
\includegraphics{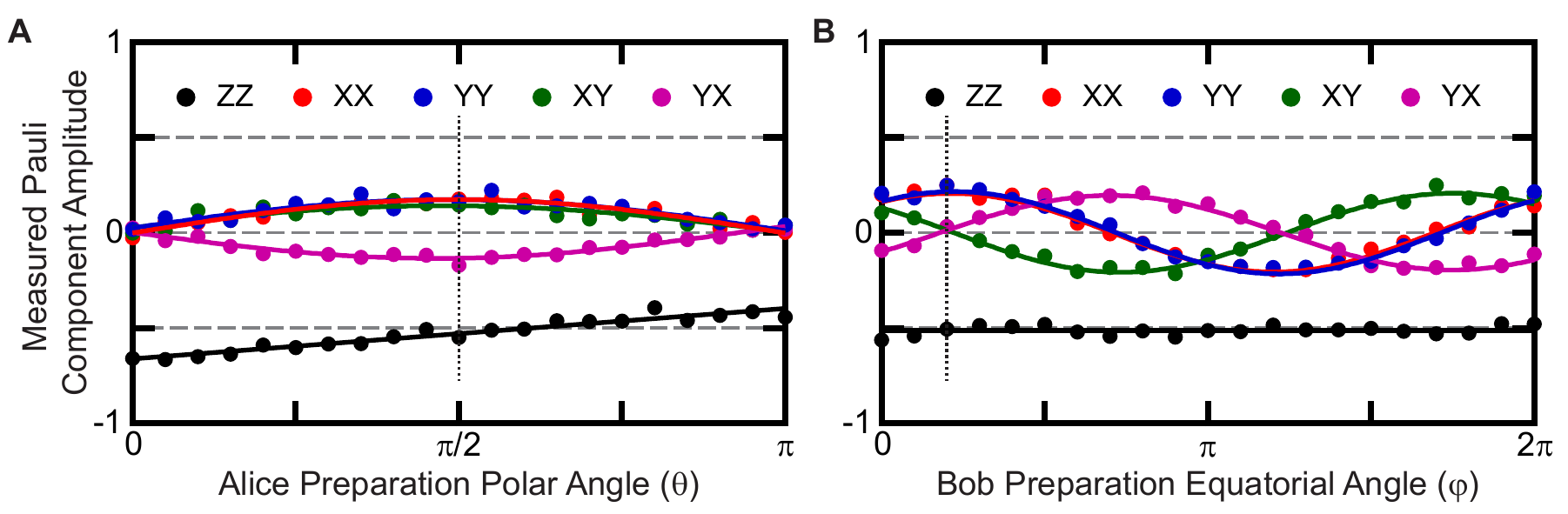}
\caption{\textbf{$\mid$ Two-qubit remote entanglement.}
Measured amplitudes of the relevant two-qubit Pauli vector components as a function of qubit preparation. After the remote entanglement protocol described in Fig.~1B, joint tomography was performed on the qubits conditioned on the detector reporting a click for each round.
A) With Bob always initialized in $\frac{1}{\sqrt{2}}\left(\g + \e\right)$, Alice was prepared in the variable state $\cos\left(\theta/2\right)\g + \sin\left(\theta/2\right)\e$. Data (points) and fits (lines) confirm that entanglement is maximized when $\theta=\pi/2$ (dotted line).
B) With Alice always initialized in $\frac{1}{\sqrt{2}}\left(\g + \e\right)$, Bob was prepared in the variable state $\frac{1}{\sqrt{2}}\left(\g + e^{i\phi}\e\right)$. The components of the Pauli vector oscillate with $\phi$ sinusoidally as expected. The complete density matrix for $\phi$ given by the dotted line is shown in Fig.~4 (left) in the Pauli basis.}
\end{figure*}

\section{Experimental Results}

As a preliminary step towards the realization of the full remote entanglement protocol, we demonstrate in Fig.~2B and C signatures of entanglement between the Alice qubit and its corresponding traveling photon state by showing that the CNOT-like operation maps $\alpha\ket{g0} + \beta\ket{e0}$ to $\alpha\ket{g0} + \beta\ket{e1}$ (for data on the Bob qubit and simulations, see Appendix C). We first show that the relative weights of $\g$ and $\e$ were correctly mapped by initializing the qubit in $\cos\left(\theta/2\right)\g + \sin\left(\theta/2\right)\e$, followed by a $\pi$~pulse on the $\omega_{A,ef}^0$ and sideband pulses for a varying time $T_{\textrm{SB}}$ (see Fig.~2A, right). The selective $\pi$-pulse on the detector was a $480$~ns Gaussian pulse ($\sigma=120$~ns) and was timed such that the center of the Gaussian coincides with the end of the sideband pulse. Finally, we measured the probability of detecting a photon in the detector, $P_{\textrm{click}}$, and the Alice polarization, $\bket{Z_A}$. As shown in Fig.~2B (black dashed line), a $\pi$-pulse on the $\ket{f0} \leftrightarrow \ket{e1}$ transition occurs for $T_{\textrm{SB}}=254$~ns when the probability of detecting a photon, $P_{\textrm{click}}$ is maximized. On the other hand, for shorter sideband pulse lengths, no photons are generated and $P_{\textrm{click}}=0$. Moreover, the observed increase in $P_{\textrm{click}}$ with $\theta$ confirms that the relative weight of the superposition state between $\g$ and $\e$ is mapped onto $\ket{g0}$ and $\ket{e1}$ (Fig.~2B, left). We also confirm that this process does not destroy the qubit state by observing that the final value of $\bket{Z_A}$ agrees with the initial preparation angle $\theta$ (Fig.~2B, right). 

Furthermore, in Fig.~2C, we show that this operation also maps the phase of $\alpha\g + \beta\e$ onto $\alpha\ket{g0} + \beta\ket{e1}$. Directly measuring the phase of $\ket{e1}$ relative to $\ket{g0}$ is not possible in this experiment since the detector only detects the presence or absence of a photon. Instead, the Alice qubit was first prepared on the equator of the Bloch sphere in $\frac{1}{\sqrt{2}}\left(\g + e^{i\phi}\e\right)$, the CNOT-like operation was either performed or not and finally both $P_{\textrm{click}}$, and the qubit equatorial Bloch vector components, $\bket{X_A}$ and $\bket{Y_A}$, were measured. When no photon is generated, $P_{\textrm{click}}=0$ as expected and $\bket{X_A}$ and $\bket{Y_A}$ oscillate with the preparation phase $\phi$ (Fig.~2C, top). However, when the operation is performed, a photon is generated and thus $P_{\textrm{click}}$ is now non-zero. Since, the preparation phase $\phi$ is now mapped onto the entangled state, the measurement of the photon, either by the detector or some other loss in the system, results in the unconditional dephasing of the qubit, $\bket{X_A}, \bket{Y_A}=0$ (Fig.~2B, bottom).

\begin{figure*}[!ht]
\includegraphics{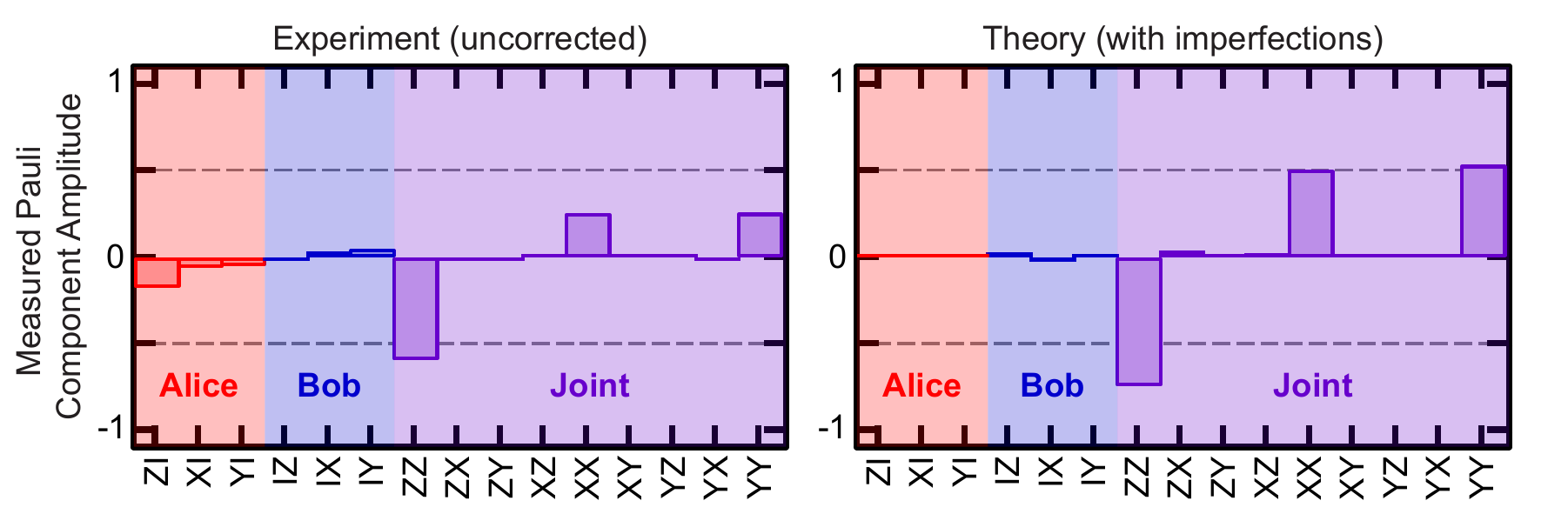}
\caption{\textbf{$\mid$ Entanglement characterization.} 
Left: Experimentally measured Pauli vector components of the two-qubit entangled state confirming that the final state is the odd Bell state $\frac{1}{\sqrt{2}}\left(\ket{ge}+\ket{eg}\right)$ with raw fidelity $\mathscr{F}=0.53$. Right: The theoretically expected Pauli components accounting for qubit decoherence, detector dark counts and tomography infidelity.}
\end{figure*}

Having demonstrated qubit-photon entanglement, we next perform the full remote entanglement protocol. The final two-qubit density matrix was measured in the Pauli basis with joint tomography (see Appendix B) conditioned on detecting two clicks. For an arbitrary Bell state, the only non-zero Pauli components are $\bket{ZZ}$, $\bket{XX}$, $\bket{YY}$, $\bket{XY}$ and $\bket{YZ}$, which are displayed in Fig.~3. We first demonstrate that the protocol entangles the qubits only when they start in the correct state. With Bob initialized in $\frac{1}{\sqrt{2}}\left(\g + \e\right)$, Alice was prepared in $\cos\left(\theta/2\right)\g + \sin\left(\theta/2\right)\e$. Entanglement is maximized for $\theta=\pi/2$ (see Fig.~3A dotted line), with extremal values for $\bket{XX}$, $\bket{YY}$, $\bket{XY}$ and $\bket{YX}$, and with the expected negative $\bket{ZZ}$ indicating a state of odd parity. On the other hand, for $\theta=0$ $\left(\theta=\pi\right)$, the final two-qubit state should be the separable state $\ket{eg}$ $\left(\ket{ge}\right)$ as indicated by $\bket{XX}=\bket{YY}=\bket{XY}=\bket{YX}=0$ and $\bket{ZZ}<0$. We attribute the deviation of $\bket{ZZ}$ from $-1$ to the dark counts in the detector and the finite $T_1$'s of the two qubits.

Next, we show that when both qubits are initialized along the equator of the Bloch sphere, remote entanglement is always generated. Alice was now prepared in $\frac{1}{\sqrt{2}}\left(\g + \e\right)$ with Bob prepared in $\frac{1}{\sqrt{2}}\left(\g + e^{i\phi}\e\right)$. In this case, the final state should be $\frac{1}{\sqrt{2}}\left(\ket{ge}+e^{i\left(\phi+\phi_{\textrm{off}}\right)}\ket{eg}\right)$, where $\phi_{\textrm{off}}$ is an arbitrary offset phase included to account for frequency offsets and path length differences between the two flying photons. This Bell state is witnessed by the tomography results in Fig.~3B, where $\bket{ZZ}$ is constant and negative while the other four displayed Pauli components follow the expected sinusoidal behavior. From the fits to the data, we extract $\phi_\textrm{off}=3\pi/10$.

The complete density matrix, $\rho_{\textrm{meas}}$, is shown in Fig.~4 (left) in the Pauli basis for $\phi$ given by the dotted line in Fig.~3B, where the fidelity $\mathscr{F}=\textrm{Tr}\left(\rho_{\textrm{meas}}\ket{O^+}\bra{O^+}\right)$ is maximum. The theoretically calculated density matrix, (Fig.~4, right), includes the effects of the coherence times of the Alice and Bob qubits, $T_{2\textrm{Bell}}$, the imperfections of the detector and the imperfections in the joint tomography (see Appendix B). As expected, most of the state information lies in the two-qubit Pauli components rather than the single-qubit ones. The measured fidelity $\mathscr{F}=0.53\pm0.01$ and concurrence $\mathscr{C}=0.1\pm0.01$\cite{Wootters1998} exceed the entanglement threshold. The error bars for the fidelity and concurrence were determined by the statistical noise from the number of measurements used for each tomography axis (see Appendix F). When accounting for systematic errors in tomography (see Appendix B), we obtain the corrected fidelity $\mathscr{F_\textrm{corr}}=0.57\pm0.01$. This fidelity can be understood as a result of various imperfections in the entanglement generation protocol: (1) decoherence of the two qubits which limits the fidelity to $\mathscr{F}_{T_{2\textrm{Bell}}}$ and (2) imperfections of the detector which are characterized by $\mathscr{F}_{\textrm{det}}$. From the measured value of $T_{2\textrm{Bell}} = 6~\mu$s and the protocol time, $T_{\textrm{seq}}=2.5~\mu$s, we expect $\mathscr{F}_{T_{2\textrm{Bell}}}\cong0.8$. The infidelity associated with the imperfect detector is characterized by the dark count ratio $P_{\textrm{d}}/P_{\textrm{click}}$, which is the fraction of detection events that are reported as clicks even though no actual photon was sent. In this experiment, $P_{\textrm{d}}/P_{\textrm{click}}=0.05$, primarily limited by the finite selectivity of the detection pulse and the imperfect readout of the detector qubit, which results in $\mathscr{F}_{\textrm{det}}\cong0.9$. A theoretical model incorporating these two imperfections was used to calculate an expected fidelity $\mathscr{F_\textrm{thy}}=0.76$ (see Appendix F). The remaining infidelity is a result of sources that are harder to characterize and will need to be explored in further work, like, for instance, the imperfections of the CNOT-like operation and the distinguishability of the photons. Nevertheless, the current results clearly establish the viability of this protocol and, by extension, the modular architecture for superconducting qubits.

Another figure of merit for this experiment is the entanglement generation rate which is determined by the repetition rate, $T_{\textrm{rep}} = 21~\mu$s, and the success probability of the experiment. The latter is determined by the product of state initialization via post-selection ($57\%$) and the detector click probability in the first ($8\%$) and second ($9\%$) rounds respectively leading to an overall success probability of $0.4\%$. The corresponding generation rate of about $200~\textrm{s}^{-1}$ is orders of magnitude faster than similar experiments performed with nitrogen-vacancy centers in diamond ($2\times10^{-3}~\textrm{s}^{-1}$)\cite{Bernien2013}, neutral atoms ($9\times10^{-3}~\textrm{s}^{-1}$)\cite{Hofmann2012} or trapped ion systems ($4.5~\textrm{s}^{-1}$)\cite{Hucul2014}. We note, however, that our generation rate ($200$~Hz) does not exceed the decoherence rate of the two qubits ($26$~kHz) and thus does not yet cross the threshold for fault tolerance\cite{Monroe2014,Hucul2014} though there are many prospects for enhancement.

\section{Outlook and Conclusions}

Improvements in generation rate and fidelity are possible with readily available upgrades to the hardware and software of our experiment. Firstly, a factor of $4$ increase in success can be achieved by installing the omitted second detector. Secondly, shaping the generated photons and detection pulse to mode match the flying photons to the detector would increase the detection efficiency by at least $50\%$ and hence multiply the generation rate by at least a factor of $2$. Moreover, this would reduce both the dark count fraction and the distinguishability of the traveling photons which would directly benefit the entanglement fidelity by bringing $\mathscr{F}_{\textrm{det}}$ closer to unity. Thirdly, an order of magnitude better coherence times for the two qubits have been demonstrated in similar 3D qubit-cavity systems\cite{Axline2016}, which should readily carry over to this experiment and improve $\mathscr{F}_{T_{2\textrm{Bell}}}$. Finally, the overall throughput of the experiment can be increased by an order of magnitude by the use of real-time feedback capabilities that have been recently demonstrated for superconducting qubits\cite{Liu2016, Riste2012}.

Combined, these upgrades could increase the entanglement generation rate by a few orders of magnitude to around $10$~kHz, to beyond the decoherence rates of approximately $100$~Hz experimentally demonstrated in 3D cQED-based quantum memories\cite{Reagor2016}. These 3D microwave cavity based memories can be readily integrated into the current system to store the generated remote entangled states thus allowing for the qubits to be reused to generate additional entangled pairs. Together with the ability to perform high-fidelity local operations between the qubit and the memory, this would offer the possibility of realizing remote entanglement distillation\cite{Bennett1996, Dur2003}, a crucial next step in realizing fault-tolerant modular systems.

In this work, we have demonstrated, in a single experiment, the set of tools that had been previously the exclusive privilege of quantum optics experiments: the availability of flying microwave single photon sources and detectors together with the spatial and temporal control of traveling photons to make them indistinguishable. With these tools, we have realized two-photon interference of microwave photons and the generation of loss-tolerant entanglement between distant superconducting qubits with concurrent measurements. The protocol speed and prospects for improving fidelity make this a very promising implementation for remote entanglement and the distribution of quantum information with microwave flying photons. Thus, this experiment opens new prospects for the modular approach to quantum information with superconducting circuits.

\clearpage

\setcounter{section}{0}
\renewcommand*{\thesection}{APPENDIX \Alph{section}}

\begin{widetext}
\begin{table}[h!]
\setlength{\tabcolsep}{0.5cm}
\begin{tabular} {l c c c}
	
	\hline
	Parameter & Alice & Bob & Detector\\

	\hline
	Cavity frequency $\omega_c^g/2\pi$~(GHz) & 7.6314 & 7.6316 & 7.6222\\
	Cavity bandwidth $\kappa/2\pi$~(MHz) & 0.9 & 1.2 & 0.9\\
	Qubit frequency $\omega_{ge}/2\pi$~(GHz) & 4.6968 & 4.6620 & 4.7664\\
	Anharmonicity $\alpha/2\pi$~(MHz) & 197 & 199 & 240\\
	Dispersive shift $\chi/2\pi$~(MHz) & 9 & 9 & 3\\
	$\textrm{T}_1~(\mu$s) & 140 & 85 & 90\\
	$\textrm{T}_{2, Echo}~(\mu$s) & 9 & 16 & 30\\
	
	\hline

\end{tabular}
	\caption{Alice, Bob and Detector qubit and cavity parameters}
\end{table}
\end{widetext}

\section{Experimental Setup}

	\subsection{Sample Fabrication and Parameters}
The three transmon qubits consist of Al/Al$\textrm{O}_x$/Al Josephson-junctions fabricated using a bridge-free electron-beam lithography technique\cite{Lecocq2011} on double-side-polished $3$~mm by $13$~mm chips of $c$-plane sapphire. The junctions are connected via $1~\mu$m leads to two rectangular pads ($1900~\mu$m $\times$ $145~\mu$m for Alice and Bob, $1100~\mu$m $\times$ $250~\mu$m for the detector) separated by $100~\mu$m. The qubit chips are placed in their respective rectangular indium-plated copper cavities ($21.34$~mm $\times$ $7.62$~mm $\times$ $43.18$~mm). The transmon parameters and couplings to the $\textrm{TE}_{101}$ cavity mode were designed using using finite-element simulations and black-box quantization\cite{Nigg2012}. Experimentally measured device parameters are listed in Table 1. 

A coaxial coupler was used as the input port of each cavity with the length of pin determining the input coupling quality factor $Q_\textrm{in}\sim10^6$. The output port for each cavity was an aperture in the cavity wall at the anti-node of the $\textrm{TE}_{101}$ mode. The size of the aperture was chosen so that $Q_\textrm{out}=7.5\times10^3$ yielding a total cavity bandwidth $\kappa\simeq1/Q_\textrm{out}$. Waveguide to coaxial cable adapters (WR-102 to SMA) were used on the output port of the cavities; since the qubit frequency is below the cutoff frequency of the waveguide while the cavity frequency is inside the passband, this section of waveguide acts as a Purcell filter for the qubit.

As shown in Fig.~5, the qubits were mounted to the base stage of a cryogen-free dilution fridge maintained below $50$~mK. The cavities were housed inside $\mu$-metal (Amumetal A4K) cans to shield them from magnetic fields. The input and output lines connected to the experiment were filtered with home-made lossy Eccosorb filters, commercial low-pass microwave filters, attenuators and isolators to attenuate radiation incident on the experiment. A commercial cryogenic HEMT amplifier was used at $3$~K to additionally amplify the output signals before subsequent room-temperature amplification and demodulation. 

\begin{figure*}
\includegraphics{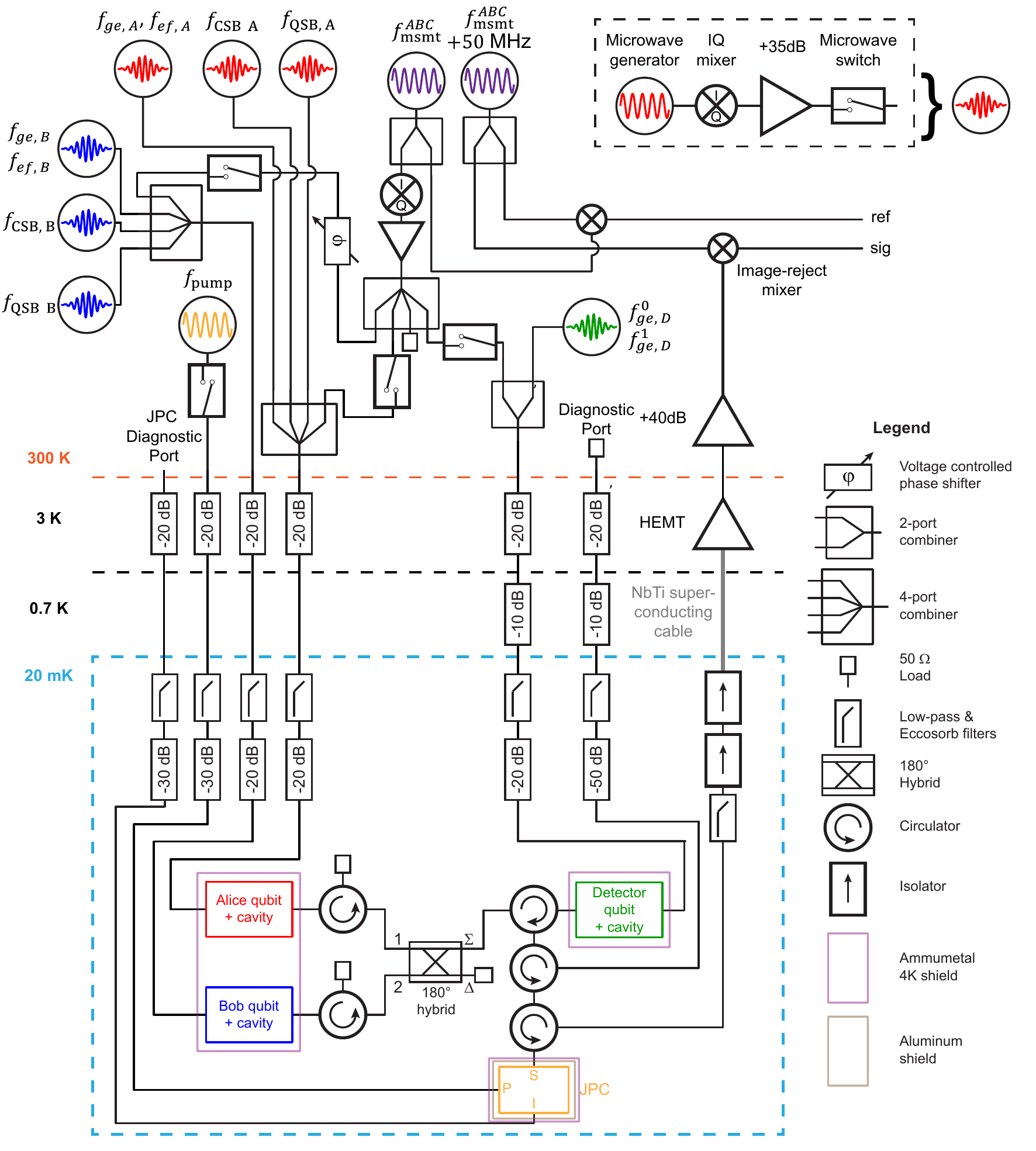}
\caption{\textbf{$\mid$ Detailed experimental setup.} 
The experiment (bottom) was cooled down on the base-stage ($<50$~mK) of a dilution refridgerator. Input lines carrying signals to the systems were attenuated and filtered using commercial low-pass filters and homemade lossy Eccosorb filters. The room temperature electronics used to produce and shape the input signals are shown at the top of the figure. The basic setup to produce shaped signals was a microwave generator driving an IQ mixer followed by an amplifier and finally a switch to gate the signal (box in top right corner). The signals were shaped by channels from four Arbitrary Waveform Generators (AWGs) (not shown) which also provided the digital markers for the switches. Copies of this setup (denoted by the shorthand notation of a circle with a shaped pulse) were used to generate the drive signals (color-coded) for three modules, Alice (red), Bob (blue) and the detector (green). The Alice and Bob modules had $4$ inputs each, the cavity readout tone, the qubit signals and the pair of sideband pulses for photon generation. On the other hand, the detector module had $2$ inputs, the cavity readout tone and the qubit signals. All the modules were readout using a single output line that had multiple stages of amplification. High-fidelity single-shot readout was enabled by the JPC amplifier. The output signals were downconverted and then digitized and demodulated along with a room-temperature reference copy.}
\end{figure*}

A critical requirement for the experiment was matching the frequencies of the Alice and Bob cavities to render the flying single photons indinstinguishable. In addition, the detector cavity frequency needs to also be matched to the Alice and Bob cavity frequencies so that incident photons can enter the detector cavity. This was achieved by an aluminum screw inserted into each cavity at the $\textrm{TE}_{101}$ anti-node to fine-tune the cavity frequencies until they satisfied $\omega_{A}^e=\omega_{B}^e=\omega_{D}^g$ (see Fig.~6A). 

\begin{figure*}
\includegraphics{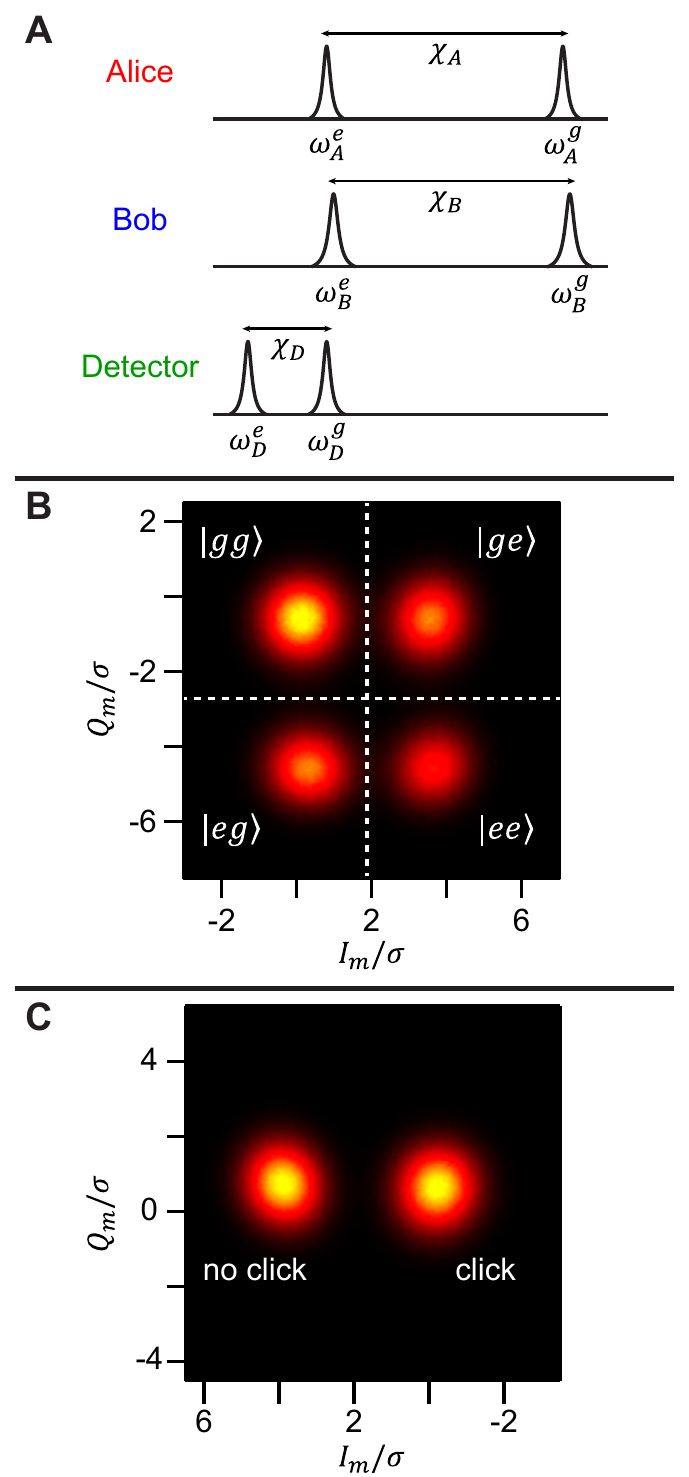}
\caption{\textbf{$\mid$ Alice, Bob and Detector qubit readout spectra and histograms.}
A) Alice, Bob and Detector cavity frequency spectra. The Alice and Bob cavities had nearly identical frequencies ($\omega_{A}^g\approx\omega_{B}^g$) and dispersive shifts ($\chi_A\approx\chi_B$). To perform joint readout of Alice and Bob, microwave pulses were simultaneously applied on each cavity at $\omega_{A}^g$ with a relative phase of $\pi/2$ between the two pulses. The detector module cavity frequency $\omega_{D}^g$ was tuned to match the frequency of the photons in the experiment, $\omega_{A}^e$. The detector was readout at $\omega_{D}^g$.
B) Joint readout histogram for Alice and Bob. A $2~\mu$s measurement pulse was used to measure the state of both qubits. The resulting output contained information about the state of Alice and Bob along the $Q_m$ and $I_m$ axes respectively. Thus, the measurement provided single-shot readout of both qubit states as well as the correlation between the two qubit states with $\mathscr{F}_{\textrm{joint}}>90\%$.
C) Readout histogram for the detector. The state of the detector qubit was measured with $\mathscr{F}_{\textrm{det}}>99\%$ in $700$~ns.}
\end{figure*}

	\subsection{Readout}
All three qubit-cavity systems were measured on the same output line using a single Josephson Parametric Converter (JPC) operated as a nearly-quantum-limited phase-preserving amplifier. The JPC was biased to provide $20$~dB of gain with a bandwidth of $8$~MHz centered at $7.6314$~GHz to realize high-fidelity single-shot readout of all three qubit-cavity system. At this operating point, a noise visibility ratio (NVR)\cite{Narla2014} of $8$~dB was measured, indicating that $86\%$ of the noise measured at room temperature was amplified quantum fluctuations from the JPC. 

As shown in Fig.~5, readout pulses for the three cavities were generated using a single microwave generator powering an IQ-mixer. The output of the mixer was split and sent to each cavity on separate input lines with the relative room temperature attenuation on each line adjusted so that an applied readout amplitude at room temperature resulted in the same measured signal-to-noise ratio (SNR) for each qubit-cavity system. Room temperature microwave switches were used on each line to gate the pulses generated by the IQ-mixer. The amplified cavity outputs were mixed down to radio frequencies along with a copy of the generator tone that did not pass through the cryostat to provide a reference. The signal and reference were digitized and demodulated to yield in-phase and quadrature components ($I(t)$, $Q(t)$) that are insensitive to drifts in the generator and other microwave components. With this setup, high-fidelity readout of all the modules in the fridge was possible with minimal hardware and complexity. In the experiments described in this paper, two types of measurements were performed: (1) joint measurement of the Alice and Bob qubits and (2) single qubit measurement of the detector 

	\subsubsection{Joint Alice and Bob measurement}

The Alice and Bob cavities were measured jointly by energizing them with $2~\mu$s pulses at $f_{msmt}^{ABC}=\omega_{A}^g/2\pi=7.6314$~GHz. Using a phase shifter on the Bob cavity arm, the relative phase of the pulses on the Alice and Bob cavities (including all system path lengths) was adjusted to $\pi/2$. The output signals from each cavity then passed through the hybrid whose output was the sum of the two cavity signals but with half the power from each signal was lost in the cold $50~\Omega$ load. This joint output signal reflects off the detector cavity (since it is $\chi_A$ above $\omega_{D}^g$) and was amplified by the JPC. As a result, the output signal demodulated at $50$~MHz contained information about both qubit states along orthogonal axes (see Fig.~6B). Two separatrices (white dashed lines), the first along the $Q_m$ axis and the second along $I_m$ axis, were used to measure the state of the Alice and Bob qubits respectively. In addition, the two-qubit correlation was calculated on a shot-by-shot basis. This resulted in an overall fidelity $\mathscr{F}_{\textrm{joint}}>90\%$. A primary limitation in achieving a higher fidelity was the loss of half the information in the cold-load after the hybrid. This can be improved in future experiments by the use of a second detector and output line. While these joint tomography imperfections will ultimately impact the measured entanglement fidelity, they can be calibrated out (as we discuss later in the Joint Tomography and Calibration section).

	\subsubsection{Detector qubit measurement}

To measure the state of the detector qubit, an IF-frequency of $-9.2$~MHz was used on the IQ-mixer to generate $700$~ns pulses at $\omega_{D}^g=7.6222$~GHz. Since this is equal to $\omega_{A}^e$ and $\omega_{B}^e$, this readout is not performed simultaneously with the joint measurement of Alice and Bob described above to avoid signal interference. The amplified output from the cavity was demodulated at $59.2$~MHz resulting in the histogram shown in Fig.~6C. As explained in the main text, measuring the qubit in $\e$ corresponds to a click in the detector. In this case, the measurement fidelity, $\mathscr{F}_{\textrm{det}}>99\%$. The measurement was optimized for maximal fidelity in the shortest possible time by using a shaped pulse that minimized the cavity ring-up and ring-down time\cite{McClure2016}. Since the pulse-shape also decreased the time taken to depopulate the cavity, operations on the detector could be performed $400$~ns after the readout instead of having to wait for the natural ring-down time.

\section{Joint Tomography and Calibration}

To calculate the final state of the Alice and Bob qubit after a joint measurement, the measured in-phase and quadrature signal ($I(t)$, $Q(t)$) was converted into a digital result using two thresholds, one for Alice and one for Bob (see Fig.~6B). Since the four measured Gaussian distributions had equal standard deviations, these thresholds were straight lines equidistant from the two distributions. Thus, using the thresholds, the output voltage from each joint tomography measurement was converted into a final outcome of $\g$ or $\e$ for each qubit. By performing measurements on an ensemble of identically prepared states, these counts were converted into expectation values of the observable being measured. Fully characterizing the state of the two qubits requires measuring the $16$ components of the two-qubit density matrix. This was done in the Pauli basis using the single-qubit pre-rotations $Id$, $R_y\left(\pi/2\right)$ and $R_x\left(\pi/2\right)$ to measure the $Z$, $X$ and $Y$ components respectively of each qubit Bloch vector and the two-qubit correlators. 

However, the tomography was not perfect ($\mathscr{F}_{\textrm{joint}}\neq100\%$) and we next discuss how to understand the imperfect tomography and calibrate out its effects\cite{McKay2015, Riste2012}. The ideal joint measurement of the two-qubit state can be described using the projectors into the computational basis:

\begin{align*}
\Pi_{GG}=\left( \begin{array}{cccc}
1 & 0 & 0 & 0 \\
0 & 0 & 0 & 0 \\
0 & 0 & 0 & 0 \\
0 & 0 & 0 & 0 \\
\end{array} \right), \Pi_{GE}=\left( \begin{array}{cccc}
0 & 0 & 0 & 0 \\
0 & 1 & 0 & 0 \\
0 & 0 & 0 & 0 \\
0 & 0 & 0 & 0 \\
\end{array} \right)
\end{align*}
\begin{align*}
\Pi_{EG}=\left( \begin{array}{cccc}
0 & 0 & 0 & 0 \\
0 & 0 & 0 & 0 \\
0 & 0 & 1 & 0 \\
0 & 0 & 0 & 0 \\
\end{array} \right), \Pi_{EE}=\left( \begin{array}{cccc}
0 & 0 & 0 & 0 \\
0 & 0 & 0 & 0 \\
0 & 0 & 0 & 0 \\
0 & 0 & 0 & 1 \\
\end{array} \right)
\end{align*} 

Here, the capital letters are used to denote a measurement outcome and distinguish it from a two-qubit state. The probability of each of those $4$ outcomes is given by $p\left(j\right)=Tr\left[\Pi_j\rho\right]$ where $j=\left\{GG, GE, EG, EE\right\}$. In the case of the imperfect measurement, the state at the end of the experiment is not faithfully converted into a measurement outcome. For example, the state $\ket{gg}$ could be recorded as EG with some probability. This can be described by the $4\times4$ matrix $\bm{A}$, where $A_{ji}$ is the probability that the state $i$ is recorded as outcome $j$. Thus, four new projectors, $\Pi_j^{expt}=\Sigma_iA_{ji}\Pi_i$, can be calculated that model this imperfection. The effects of this imperfect measurement were accounted for in the theoretically calculated density matrix in Fig.~4B of the main text.

To calculate $\bm{A}$ for this system, a calibration experiment was performed where the $4$ computational states $\ket{gg}$, $\ket{ge}$, $\ket{eg}$ and $\ket{ee}$ were prepared. Then joint-tomography was performed to calculate the probability of each measurement outcome. By measuring $p_j$ for each of the input states, the values of $A_{ji}$ were calculated, yielding: 

\begin{equation}
\bm{A}=\left( \begin{array}{cccc}
0.941 & 0.047 & 0.031 & 0.001 \\
0.031 & 0.925 & 0.001 & 0.030 \\
0.027 & 0.001 & 0.931 & 0.031 \\
0.001 & 0.027 & 0.037 & 0.938 \\
\end{array} \right)
\end{equation}

With this matrix, the tomography for the actual experiment could be corrected. For a given tomography pre-rotation $k$, the outcome can be written as a vector of probabilities $B_k = \left(p\left( GG\right)_k, p\left( GE\right)_k, p\left( EG\right)_k, p\left( EE\right)_k \right)$. Thus, the experimental state in the computational basis, $P_k$, that resulted in this outcome is given by $P_k=\bm{A}^{-1}B_k$. This operation was applied to tomography outcomes to calculate a corrected density matrix, $\rho_{\textrm{corr}}$, and thus a corrected fidelity, $\mathscr{F}_{\textrm{corr}}=57\%$.

\section{Qubit-Photon Entanglement}

\begin{figure*}
\includegraphics{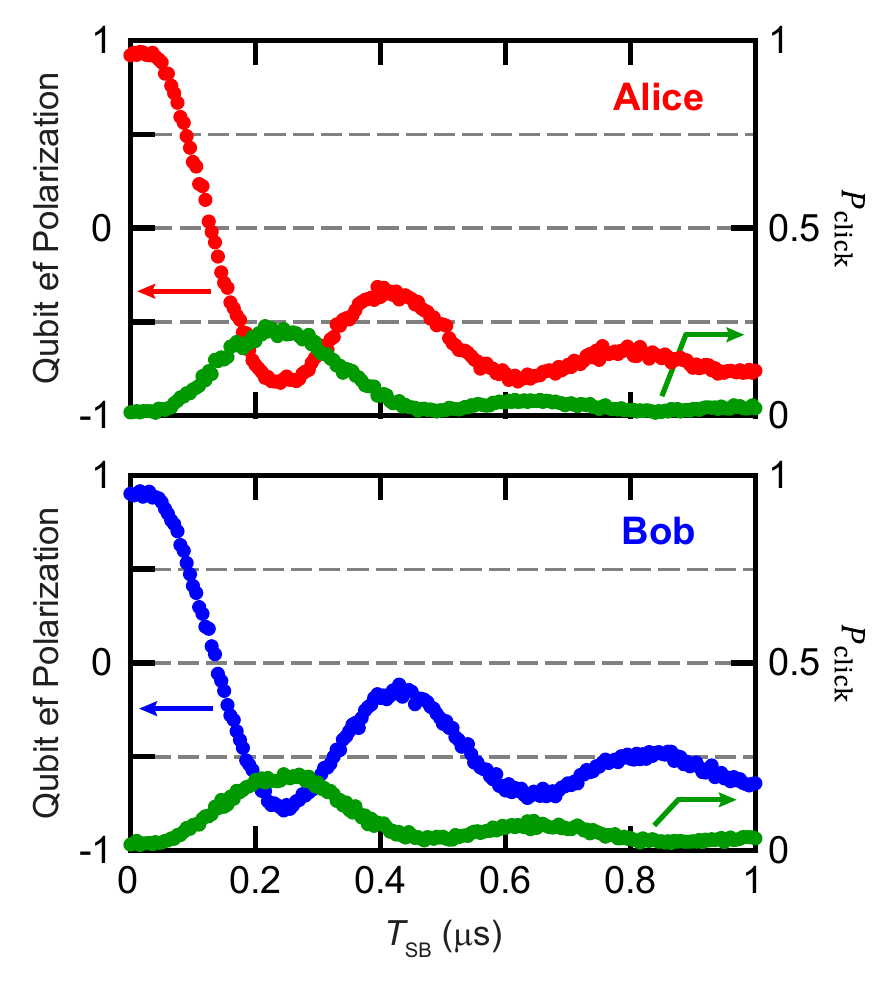}
\caption{\textbf{$\mid$ Single photon generation with sideband transitions.} 
Alice (top) and Bob (bottom) qubit $ef$ polarization (left axis) and detector click probability, $P_{\textrm{click}}$, (right axis) as a function of sideband pulse length, $T_{\textrm{SB}}$, when the qubit is prepared in $\f$. Two sideband drives ($\omega_{QSB}$, $\omega_{CSB}$) were applied, satisfying the frequency condition $\omega_{CSB}-\omega_{QSB}=\omega_{A/B}^e-\omega_{ef}$. The drives result in coherent oscillations between $\ket{f0}$ and $\ket{e1}$ with the amplitude of the drives chosen that a $\pi$-pulse on the transition took the same time for the Alice and Bob qubits, $T_{\textrm{SB}}=254$~ns. The generation of a photon was verified with the detector which showed a peak in $P_{\textrm{click}}$ when Alice/Bob were in $\e$.}
\end{figure*}

As discussed in the main text, the CNOT-like operation that entangles the stationary qubits with flying microwave photons is realized by a $\pi$-pulse on the qubit $ef$-transition followed by a $\pi$-pulse between $\ket{f0} \leftrightarrow \ket{e1}$ following the method in Ref. \cite{Wallraff2004, Kindel2015}. To drive coherent transitions between $\ket{f0} \leftrightarrow \ket{e1}$, two sideband tones at $\omega_{QSB, A}/2\pi=5.1987$ and $\omega_{CSB, A}/2\pi=8.3325$ ($\omega_{QSB, B}/2\pi=4.9631$ and $\omega_{CSB, B}/2\pi=8.1302$) were applied to Alice (Bob). As shown in Fig.~7, these drives result in damped sideband Rabi oscillations of the qubit state between $\f$ and $\e$ (Alice top, Bob bottom).  The probability of detecting a photon with the detector, $P_{\textrm{click}}$,  shown on the right axes of the graphs in Fig.~7, peaked when the qubit was in $\e$ confirming that a photon is generated. Thus, a $\pi$ pulse can be performed by turning on the drives for half an oscillation, i.e. the time taken to transfer the excitation from the qubit to the cavity. The amplitudes of the CSB and QSB drives on Alice and Bob were chosen so that the $\pi$-pulse on $\ket{f0} \leftrightarrow \ket{e1}$ took the same time, $T_{\textrm{SB}}=254$~ns, for both modules. While the oscillations would ideally be between $+1$ and $-1$, a deviation from this behavior is observed in the data. We attribute this behavior to the QSB tone spuriously exciting the $ge$ and $ef$ transitions and hence driving the qubit out of $\e$. While increasing the detuning of the drives would lower the spurious excitation, this was not possible in our experiment because of power limitations. Similarly, the drive amplitudes could have been decreased but this would have increased the photon generation time and degraded the fidelity of two-qubit entangled state because of decoherence. Thus, the drive amplitudes and detunings were chosen to balance the two effects.

Using the CNOT-like operation, signatures of qubit-photon entanglement for the Alice module were demonstrated in Fig.~2 of the main text. Similar signatures were observed for the Bob module as shown in Fig.~8. The observed behavior agrees with the results of a simplified theoretical model (right panels, Fig.~8A). In this model, the action of the sideband drives on Alice/Bob was modeled using the theory of damped vacuum Rabi oscillations described in \cite{Haroche2006}. We note that although our system uses sideband transitions between a different set of states, the coupling can still be modeled with the same formalism. Thus, the three states used here were $\ket{f0}$, $\ket{e1}$ and $\ket{e0}$. The sidebands drive coherent transitions between $\ket{f0}$ and $\ket{e1}$ while the cavity linewidth, $\kappa$, causes $\ket{e1}$ to decay to $\ket{e0}$. For the detector signal, we made the simplification of using the state of the cavity subjected to two inefficiencies as a proxy. Thus, $P_{\textrm{click}}\left(T_{\textrm{SB}}\right)=\eta P_{e1}\left(T_{\textrm{SB}}\right)$, where $\eta$ accounts for the loss between the Alice/Bob module and the detector as well as the detector efficiency and $P_{e1}\left(t\right)$ is the probability of the system being in $\ket{e1}$. We find good qualitative agreement between theory and experiment.

\begin{figure*}
\includegraphics{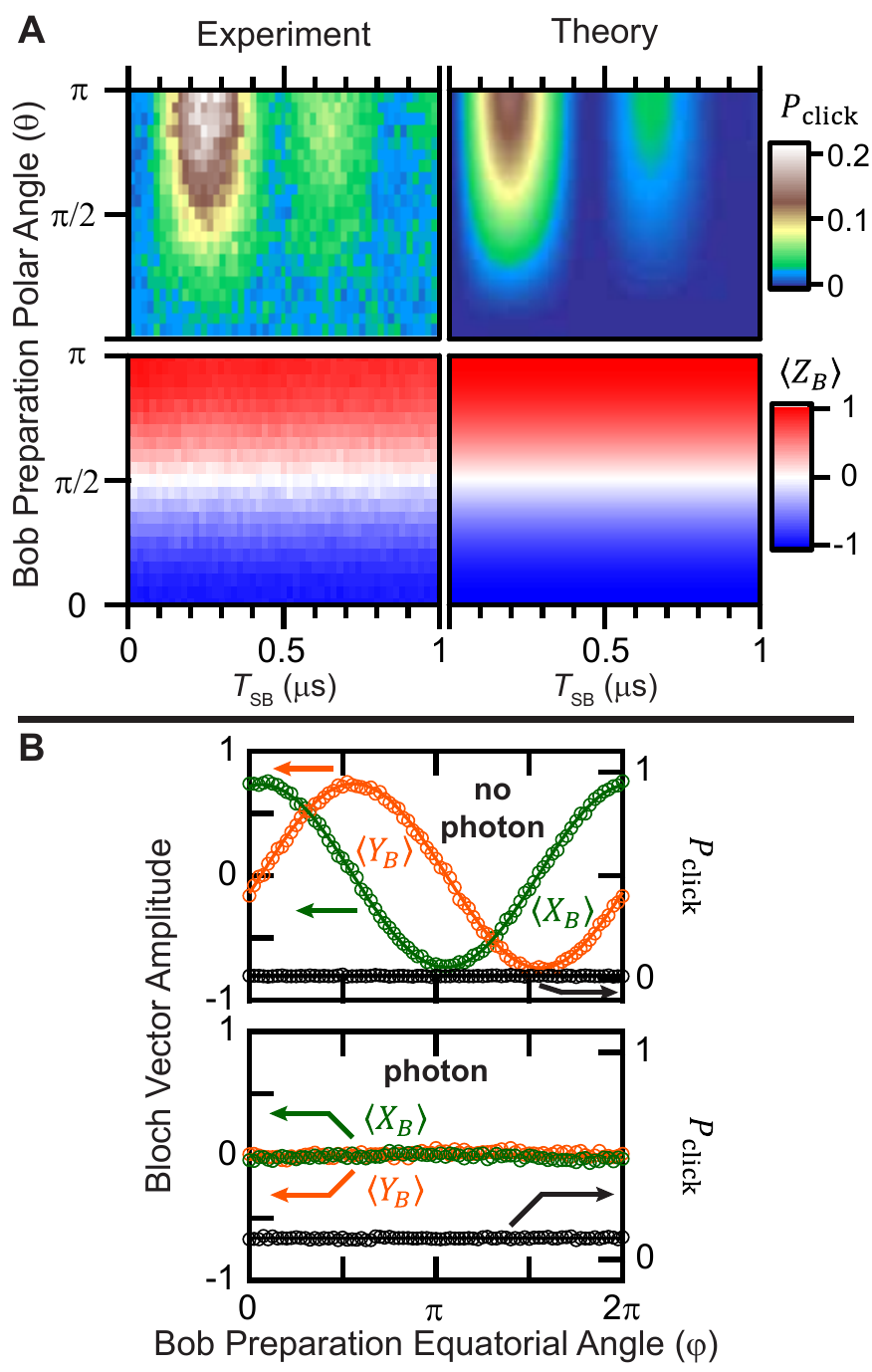}
\caption{\textbf{$\mid$ Signatures of Qubit-Photon entanglement: data vs. theory.} 
This data is similar to that of Fig. 2 but addresses the Bob module of the experiment and its equivalence to the Alice module.
A) Color plots of the probability, $P_{\textrm{click}}$ (top left), of the detector qubit ending in $\e$ and the Bob qubit polarization, $\bket{Z_B}$ (bottom left), as a function of the sideband pulse length $T_{\textrm{SB}}$ when Bob was prepared in $\cos\left(\theta/2\right)\g + \sin\left(\theta/2\right)\e$. A theoretical simulation, plotted on the right, shows good agreement. 
B) Detector click probability, $P_{\textrm{click}}$, and Bob equatorial Bloch vector components, $\bket{X_A}$ and $\bket{Y_A}$, as a function of $\phi$ when  Bob was prepared in $\frac{1}{\sqrt{2}}\left(\g + e^{i\phi}\e\right)$ and the CNOT-like operation was either performed (bottom) or not (top). Open circles are experimental data and lines are fits.}
\end{figure*}

\section{Microwave Photon Detector}

	\subsection{Simulations} 

\begin{figure*}
\includegraphics{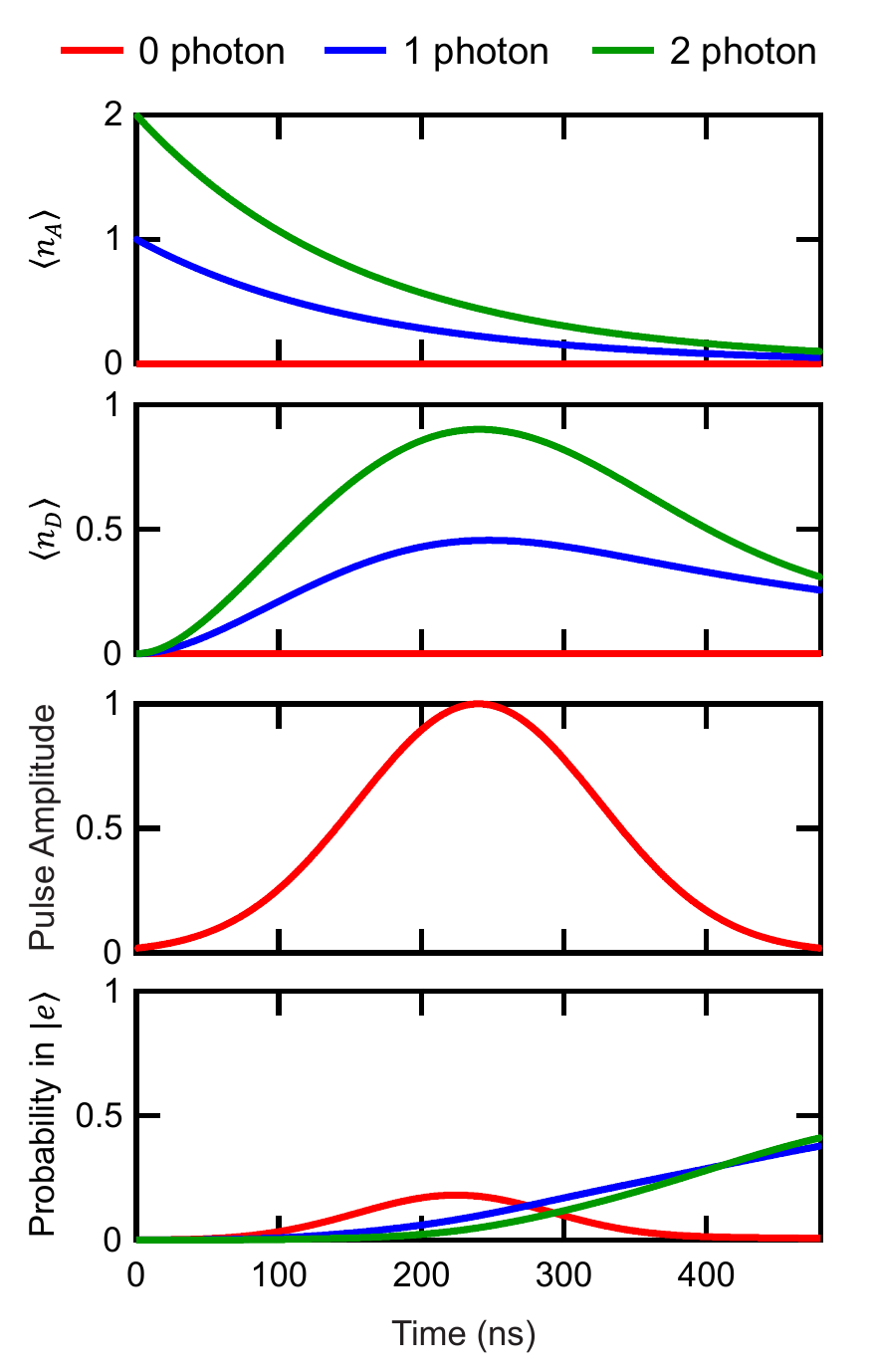}
\caption{\textbf{$\mid$ Detector Simulations.} 
Results from solving the master equation for a cascaded quantum system of the Alice cavity emitting Fock states into the detector qubit-cavity system. The top two panels show the expectation value of the photon number operators of the Alice, $\bket{n_A}$, and detector, $\bket{n_D}$, cavities. The Alice cavity (top panel) was initialized in $\ket{0}$ (red trace), $\ket{1}$ (blue trace) or $\ket{2}$ (green trace). The third panel shows the amplitude of a selective $\pi$-pulse with $\sigma=120$~ns applied on the detector qubit to excite it conditioned on the presence of a singe intra-cavity photon. Finally, the probability to find the detector qubit in $\e$ was calculated to find $P_{\textrm{click}}$ at the end of process (bottom panel). Simulations confirm that the detector has dark counts ($P_{\textrm{click}}$ given $\ket{0}$) $P_{\textrm{d}}<0.01$ and an efficiency ($P_{\textrm{click}}$ given $\ket{1}$) $\eta\sim0.4$. Since $P_{\textrm{click}}$ is the same for $\ket{1}$ (blue trace) and $\ket{2}$ (green trace), the detector is not number-resolving.}
\end{figure*}

A cascaded quantum system simulation\cite{Carmichael1993, Gardiner2004, Fan2014} was performed to understand the operation of the detector and how two characteristics, dark counts and detector efficiency, depend on system parameters. We simulate a simplified model of the experiment consisting of a single emitter cavity, Alice, and the detector qubit-cavity module. The master equation for this system was solved for various initial states of Alice modeling the inputs seen by the detector in the experiment. The simulations were performed with the experimentally measured parameters (see Table 1). However, unlike the experiment, the two cavities had identical cavity frequencies.

As shown in Fig.~9, the simulation began by initializing the Alice cavity in the $\ket{0}$ (red trace), $\ket{1}$ (blue trace) or $\ket{2}$ (green trace) Fock state (top panel). The photon leaked out and excited the detector cavity (second panel). Simultaneously, a selective $\pi$-pulse, timed to start at the beginning of the simulation, with $\sigma=120$~ns was applied at $\omega_{ge}^1$ to selectively excite the detector qubit conditioned on the presence of a intra-cavity photon (third panel). Finally, $P_\textrm{click}$ was extracted by calculating the probability that the detector qubit state was $\e$ at the end of the simulation (bottom panel). The first detector characteristic, its dark count fraction $P_\textrm{d}$, is the probability that the detector clicks when the input is $\ket{0}$. When no photons were sent to the detector (red trace), $P_\textrm{click}<0.01$ at the end of the simulation. The transient increase in the probability of the detector qubit being in $\e$ observed during the course of the qubit pulse is a result of the finite selectivity of the $\pi$-pulse which was confirmed by varying $\sigma$ or $\chi$. Thus, the dark count probability, $P_\textrm{d}$, can be decreased by increasing $\sigma$ at the cost of slowing down the detection process (and hence the detection probability).

The second detector characteristic is its efficiency, $\eta$, the probability that the detector clicks when the input is $\ket{1}$. When one photon was sent to the detector, the qubit was excited by the selective $\pi$-pulse resulting in $P_\textrm{click}=0.4$. On the other hand, when two photons were sent to the detector, on average a single photon entered the detector, also resulting in $P_\textrm{click}=0.4$. Since $P_\textrm{click}$ is similar for $\ket{1}$ and $\ket{2}$, the detector is not photon-number resolving. Furthermore, the simulations verified that the detector efficiency is robust to small imperfections and does not require precise tuning. When the simulation parameters, such as the mismatch between the Alice and detector cavity bandwidths and the selective pulse length and timing, were varied by $20\%$, $\eta$ changed by $<10\%$.

	\subsection{Detector Characterization}

\begin{figure*}
\includegraphics{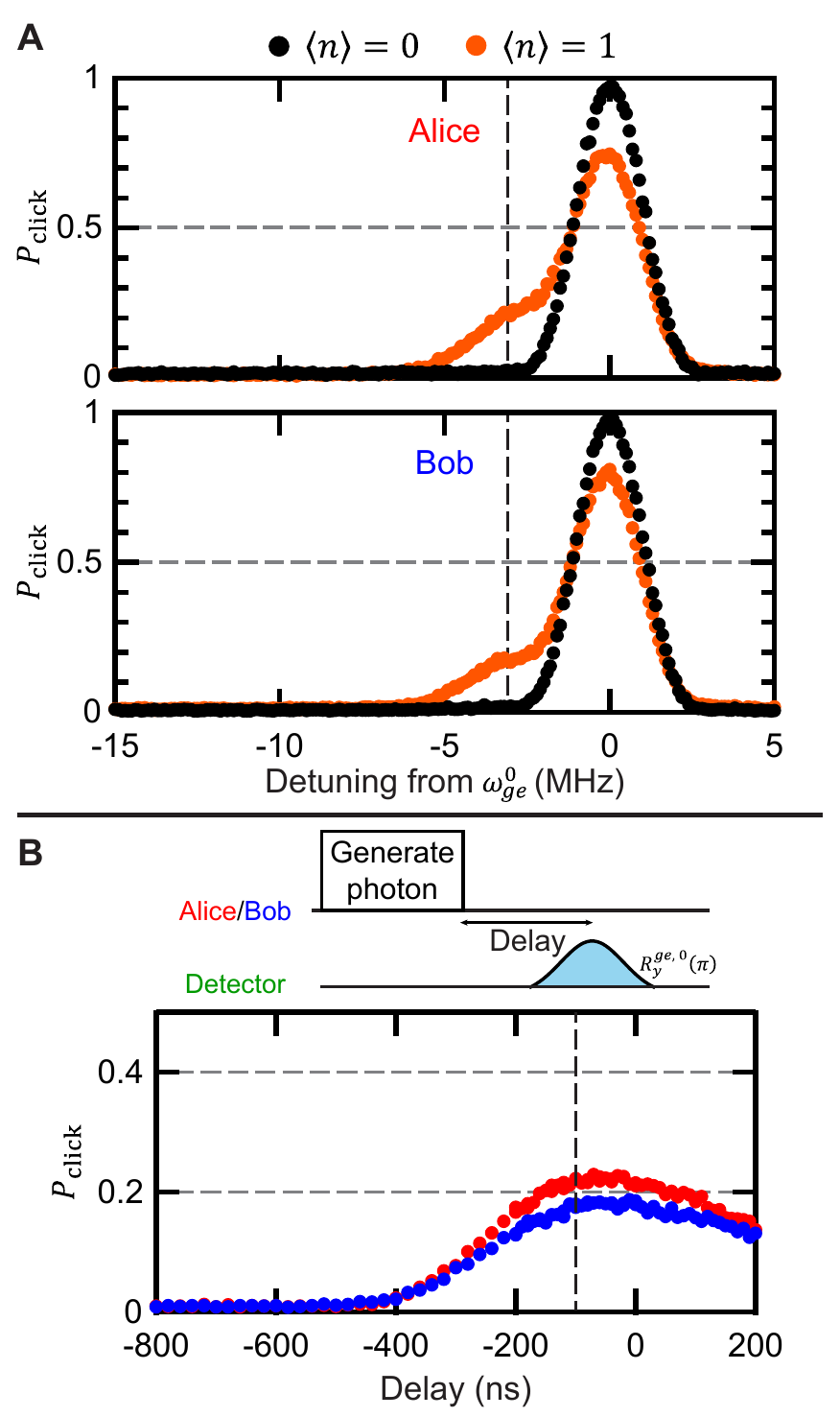}
\caption{\textbf{$\mid$ Detector characterization.} 
A) Detector click probability, $P_\textrm{click}$, as function of the detuning of the detection pulse from $\omega_{ge}^0$ for different input states from the Alice (top panel) and Bob (bottom panel) modules. The black dashed line indicated the frequency of the selective $\pi$-pulse for optimum discrimination of the state $\ket{1}$ from the state $\ket{0}$. 
B) Detector click probability, $P_\textrm{click}$, as a function of the delay between the end of the photon generation pulse and the start of the selective detection $\pi$-pulse. In the remote entanglement experiment of Figs.~3 and 4, the pulses overlapped by $100$~ns (black dashed line).}
\end{figure*}

The performance of the detector was also characterized experimentally to verify that it was detecting single photons. In these experiments (see Fig.~10A), the Alice and Bob modules were initialized in one of the two states, $\ket{0}$ or $\ket{1}$. Single photons were generated by preparing the qubit in $\e$ and then performing the CNOT-like operation to create the state $\ket{e1}$. Note that the generation process takes $254$~ns unlike the assumption of instantaneous generation in the simulations. Then, detection was performed by applying the selective $\pi$-pulse ($\sigma=120$~ns) on the detector followed by measuring the state of the detector qubit to find $P_\textrm{click}$. The frequency of the detection $\pi$-pulse was varied to characterize the detector response as a function of frequency. As shown in Fig.~10A, when the state $\ket{0}$ (blue circles) was sent, the $P_\textrm{click}$ was maximized at zero detuning where the pulse is selective on zero intra-cavity photons in the detector. Instead, when the input was $\ket{1}$ (red circles), an increased response at $\omega_{ge}^0-\chi$ was observed. This is a direct result of the detector being excited when photon enters the detector. Due to losses and the detector inefficiency, the response at zero detuning remains but with a lower $P_\textrm{click}$ than for $\ket{0}$. Moreover, the similar detector response to inputs from Alice and Bob demonstrates that the detector can detect photons from both systems and that the losses on the two arms are similar on the two paths.

\begin{figure*}
\includegraphics{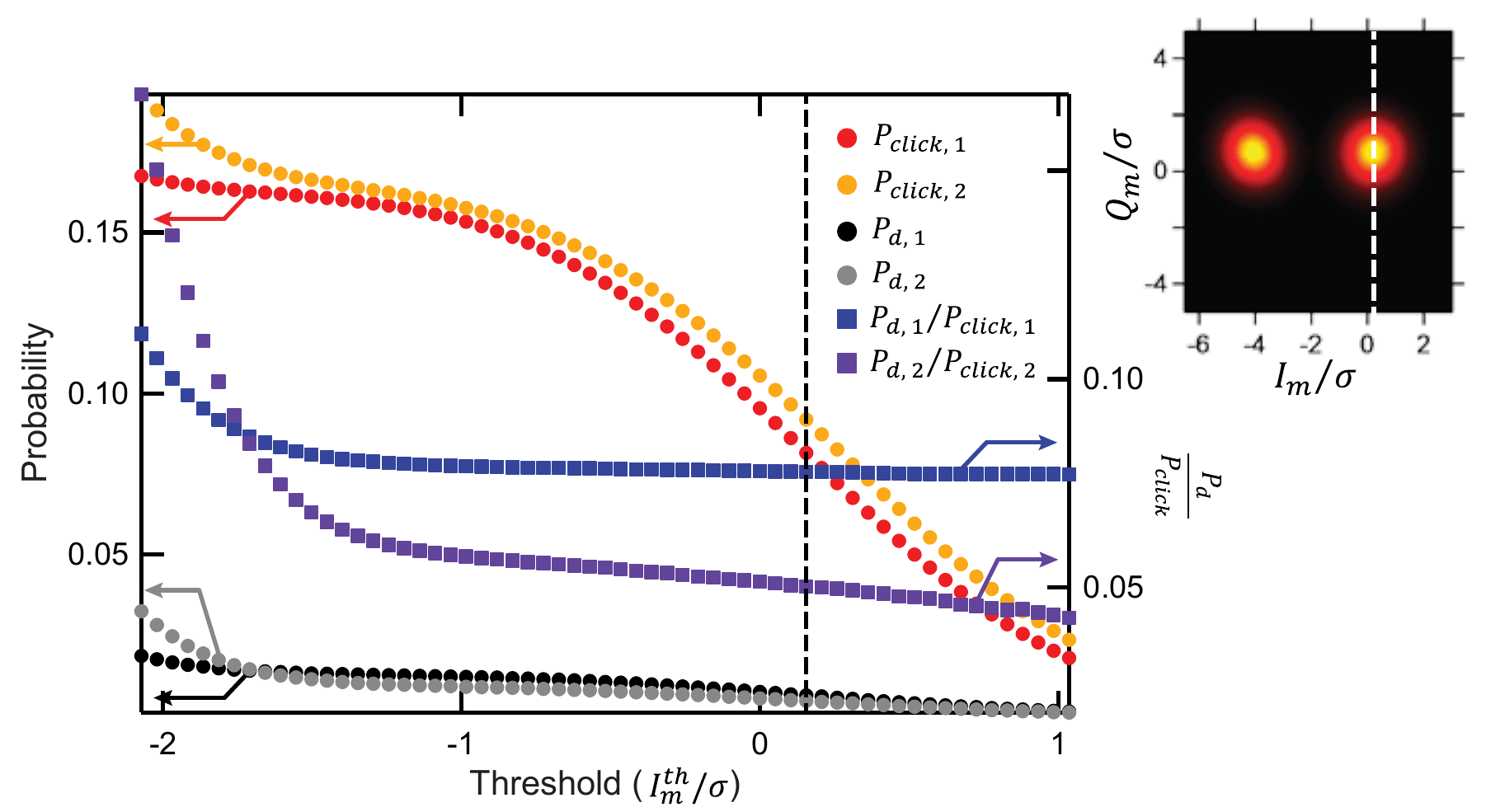}
\caption{\textbf{$\mid$ Detector optimization.} 
The probability of dark counts, $P_\textrm{d}$, and detector click probability, $P_\textrm{click}$, (left axis) and their ratio (right axis) for each round of detection as a function of the readout threshold $I_m^{th}/\sigma$. The detector readout has two probability distributions (inset), one for click and one for no click. By using a more stringent threshold for outcomes to be considered a click (white dashed line/black dashed line), the ratio $P_\textrm{d}/P_\textrm{click}$ can be reduced, therefore improving the fidelity of the generated Bell state.}
\end{figure*}

In a second characterization experiment, the delay between the end of the photon generation and beginning of the photon detection steps was optimized. The probability of detecting the photon, $P_\textrm{click}$, is maximized when the peak of the detection pulse coincides with the time at which the photon population inside the detector cavity is maximum. To find this point experimentally, a photon was generated by Alice or Bob and sent to the detector with a variable delay between the end of the photon generation sideband pulse and the beginning of the selective detection $\pi$-pulse. As shown in Fig.~10B, $P_\textrm{click}$ was maximized around a delay of $-100$~ns (black dashed line), i.e when the sideband and detection pulses had $100$~ns of overlap. This operation point was used in the remote entanglement experiments of Figs.~3 and 4.

We attribute the difference between the simulated detector efficiency, $\eta=0.4$, and the measured $P_\textrm{click}$ when a photon was generated in experiments to the losses in our system and dark counts. Due to the the hybrid and the insertion losses of the microwave components between the Alice/Bob modules and the detector, photons only reach the detector about $40\%$ of the time, corresponding to an efficiency due to the loss of $\eta_\textrm{loss}\sim0.4$. In addition, the detector can also click when no photon is incident on it, which occurred with a probability $P_\textrm{d}=0.01$. Together, they result in the observed $P_\textrm{click}\sim0.2$ when a photon was generated.

	\subsection{Detector Optimization}

This remote entanglement protocol is robust to loss since the generation of an entangled state is uniquely heralded by the dual detection of single photons in the detector. Hence, photon loss between Alice/Bob and the detector only affect the probability of that outcome. However, dark counts in the detector are detrimental to this experiment (for a quantitative discussion of the effect, see Appendix F) because they mix the desired Bell state with unwanted states, for example $\ket{gg}$. This impacts the measured fidelity. Since the desired (undesired) outcomes occur with probabilities proportional to $P_\textrm{click}$ ($P_\textrm{d}$), the ratio of $P_\textrm{d}/P_\textrm{click}$ is the figure of merit that must be minimized for reducing the infidelity due to dark counts. Thus, it is important to minimize the probability of dark counts in the detector, $P_\textrm{d}$. In our detector, dark counts occur as a result of the finite selectivity of the detection $\pi$-pulse and imperfect readout of the qubit state. While the detection pulse could be made more selective by increasing its $\sigma$, this would increase the overall detection time. Unfortunately, this has two undesired consequences. First, the overall protocol time increases, and thus, so does the infidelity due to decoherence. Second, simulations show that the detector efficiency is maximized for $\sigma\sim\kappa$ and thus increasing $\sigma$ further actually increases $P_\textrm{d}/P_\textrm{click}$. Therefore, we operated with $\sigma=120$~ns. 

\begin{figure*}
\includegraphics{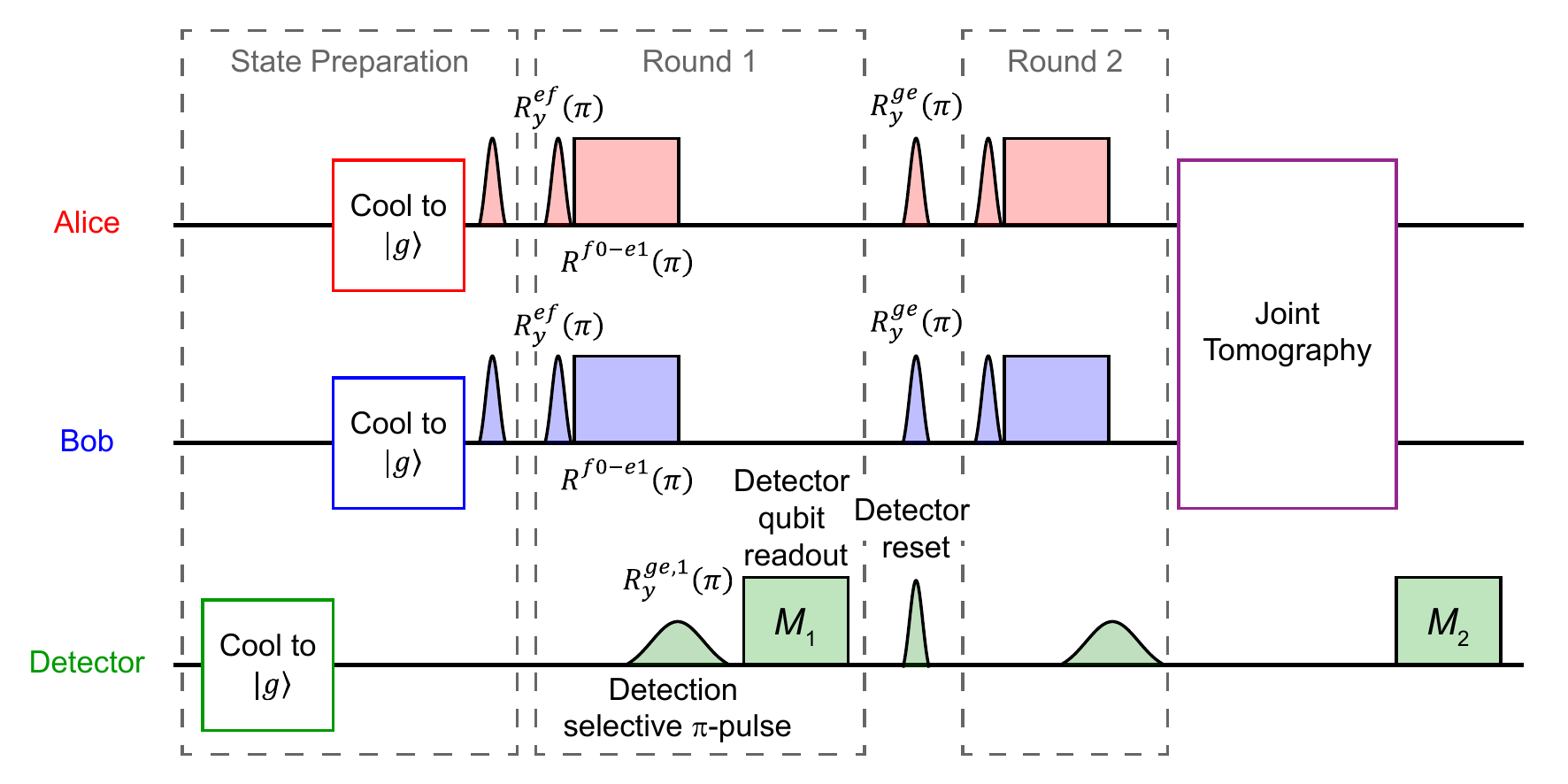}
\caption{\textbf{$\mid$ Detailed remote entanglement protocol pulse sequence.} 
The remote entanglement protocol began with state preparation where the three qubit-cavity systems were initialized in the desired state by cooling and single-qubit rotations. Then, the first of two rounds of the protocol was performed. The qubits were entangled with flying single photons by a CNOT-like operation which then interfered on the hybrid and were detected by a selective $\pi$-pulse on the detector qubit. A $\pi$-pulse was performed on both Alice and Bob to remove the unwanted $\ket{ee}$ state and the detector to reset it. Next, the second round of the protocol was performed followed by joint tomography to measure the state of Alice and Bob. The measurement outcomes from the two rounds of photon detection, $M_1$ and $M_2$, were used to post-select successful trials for the tomographic analysis. The entire protocol was repeated with $T_{\textrm{rep}}=21~\mu$s, much faster than the $T_1$ time of any of qubits.}
\end{figure*}

Instead, we decrease the ratio $P_\textrm{d}/P_\textrm{click}$ in post-selection by reducing the probability that the detector clicks when the state $\ket{0}$ is incident on it. As discussed before, readout of the detector qubit results in two distributions, one for click and one for no click. As shown in Fig.~11, by moving the threshold closer to the distribution associated with a click in the detector, it was possible to decrease the dark count fraction. The data for $P_\textrm{click}$ (red and yellow circles) and $P_\textrm{d}$ (black and grey circles) were obtained from the two rounds of the remote entanglement experiment and the control experiments (see Appendix E) respectively. From these two numbers, the ratio $P_\textrm{d}/P_\textrm{click}$ (blue and purple squares) was calculated for each round. A threshold in the middle of the two distributions corresponds to $I_m^{th}/\sigma=-1.8$ where $P_\textrm{d}/P_\textrm{click}=0.1$ for the second round. By moving the threshold to $I_m^{th}/\sigma=0.15$ (black dashed line), the ratio decreases to $P_\textrm{d}/P_\textrm{click}=0.05$.

\begin{figure*}
\includegraphics{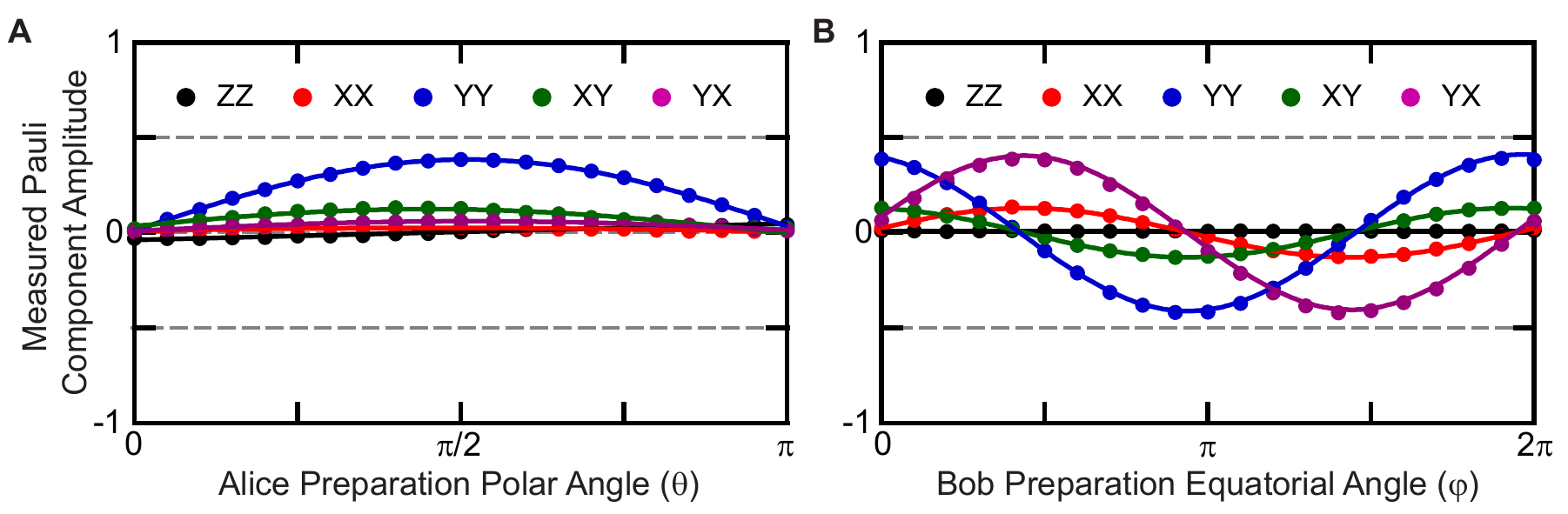}
\caption{\textbf{$\mid$ Control sequence data.} 
Measured amplitudes of selected two-qubit Pauli vector components as a function of qubit preparation. In experiments identical to those in Fig.~3, the two qubit were prepared in the desired initial state but no flying photons were generated. Joint tomography of the final two-qubit state was performed.
A) With Bob always initialized in $\frac{1}{\sqrt{2}}\left(\g + \e\right)$, Alice was prepared in the variable state $\cos\left(\theta/2\right)\g + \sin\left(\theta/2\right)\e$. 
B) With Alice always initialized in $\frac{1}{\sqrt{2}}\left(\g + \e\right)$, Bob was prepared in the variable state $\frac{1}{\sqrt{2}}\left(\g + e^{i\phi}\e\right)$. In both cases, data (points) and fits (lines) confirm that no two-qubit entanglement is observed. This is most directly indicated by $\bket{ZZ}=0$ unlike Fig.~3.}
\end{figure*}

\section{Detailed Experimental Protocol}
	
	\subsection{Pulse Sequence}
	
In the first step of the complete remote entanglement protocol (Extended Data Fig.~8), the Alice, Bob and the detector qubits were initialized in $\g$. They were first cooled to the ground states using a driven reset protocol\cite{Geerlings2013} and then a measurement was performed to post-select on experiments where all three qubits were successfully cooled. This state initialization by post-selection had a success probability of $57\%$. Moreover, this also allowed the experiment to be repeated at $T_{\textrm{rep}} = 21~\mu$s, much faster than the relaxation time of any qubit. Single qubit pulses were then applied to the Alice and Bob qubits to prepare them in the desired initial state. Then, the first round of the remote entanglement protocol consisting of the CNOT-like operation and the photon detection were performed. Before the second round, a $\pi$-pulse on $\omega_{ge}$ was applied to both the Alice and Bob qubits to remove the weight in the $\ket{ee}$ state. In addition, the detector was reset by an unselective $\pi$-pulse that returned the detector qubit to $\g$ if it went click in the first round. Such an unconditional reset can be used since only those trials where the detector went click were used in the final data analysis. After a second round of the CNOT-like operation and photon detection, joint tomography of the Alice and Bob qubit state was performed conditioned on measuring two clicks in the detector. As shown in Fig.~12, the measurement of the detector qubit in the second round was performed after the joint tomography to reduce the protocol time and hence, the effects of decoherence. This can be done because the photon detection process is completed at the end of the detection $\pi$-pulse. The measurement of the qubit state is required only for the experimenter to determine the outcome of the detection event. A set of control sequences was interleaved into the above protocol to calibrate the joint tomography . These experiments were repeated to accumulate at least $10^5$ successful shots of each sequence for adequate statistics.

	\subsection{Control Experiments}
	
To verify that the experimental results observed in the data shown in Fig.~3 are a result of the which-path erasure of the flying photons by the hybrid, two control experiments were performed. In these experiments, no flying photons were generated but the experimental protocol was otherwise left unchanged. The joint tomography performed at the end of the protocol is no longer conditioned on photon detection events. To further rule out systematic error, these experiments were interleaved with the experiments performed in Fig.~3. The results on these experiments are shown in Fig.~13. In the first experiment, a control for the data in Fig.~3A, Bob was initialized in $\frac{1}{\sqrt{2}}\left(\g + \e\right)$ and Alice was prepared in $\cos\left(\theta/2\right)\g + \sin\left(\theta/2\right)\e$. Since the qubits were not entangled with photons, no entanglement was generated for any preparation angle $\theta$. This is most directly demonstrated by $\bket{ZZ}=0$, unlike in Fig.~3A where $\bket{ZZ}<0$. Since Bob remained in $\frac{1}{\sqrt{2}}\left(\g + \e\right)$ at the end of the experiment independent of $\theta$, the final single-qubit Bloch vector has Pauli components $\bket{Z_B}=0$, $\bket{X_B}=0$ and $\bket{Y_B}=1$. Consequently, only $\bket{YY}$ and $\bket{XY}$ vary with $\theta$ and are maximized at $\theta=\pi/2$ while $\bket{XX}=\bket{YX}=0$, unlike in Fig.~3A. 

In the second experiment, a control for the data in Fig.~3B, Alice was now initialized in $\frac{1}{\sqrt{2}}\left(\g + \e\right)$ and Bob was prepared in $\frac{1}{\sqrt{2}}\left(\g + e^{i\phi}\e\right)$. In the control experiment with no photons, the final two-qubit state should be the superposition of the computation states $\frac{1}{2}\left(\ket{gg}+e^{i\phi}\ket{ge}+\ket{eg}+e^{i\phi}\ket{ee}\right)$. Thus, $\bket{ZZ}=0$ (see Fig.~13B). Moreover, $\bket{XX}$ and $\bket{YY}$ do not have in-phase sinusoidal oscillations characteristic of an odd Bell state. Ideally, $\bket{XX}=\bket{XY}=0$ but a small detuning error on the Alice qubit caused oscillations in them too. 

\section{Entanglement Fidelity}

To understand the sources of infidelity in the experiment, various sources of imperfection were built into a quantum circuit model of the entire system. The model contained both qubits, treated as two-level systems, an upper and lower branch of the experiment that could have $0$, $1$ or $2$ flying photons and two single-photon detectors. Thus, the total system state was described by a $36\times36$ density matrix. Sources of imperfections were individually introduced and their effects on this density matrix was calculated. By cascading their effects on the density matrix, their combined impact was also calculated. Finally, to compare to experiment, the photon parts of the density matrix were traced out to reduce it to a two-qubit density matrix which was expressed in the Pauli basis to generate Fig.~4B and calculate the expected fidelity. 

	\subsection{Qubit Decoherence}

The effects of qubit decoherence on the density matrix were modeled using phase damping. For a single qubit, this can be represented by the quantum operation $\mathcal{E}\left(\rho\right)=E_0\rho E_0^\dagger+E_1\rho E_1^\dagger$ \cite{Nielsen2004}. Here,

\begin{equation}
E_{0}=\sqrt{\alpha}\left( \begin{array}{cc}
1 & 0 \\
0 & 1 \\
\end{array}\right),~E_{1}=\sqrt{1-\alpha}\left( \begin{array}{cc}
1 & 0 \\
0 & -1 \\
\end{array} \right)
\end{equation} 

and $\alpha=\left(1+e^{-t/T_{\textrm{2E}}}\right)/2$. The decoherence of each qubit was treated as an independent process assuming that there was no correlated noise affecting the two systems. Thus, by taking its Kronecker product with a $2\times2$ identity matrix, the single-qubit phase damping operation was converted into a two-qubit operator. Two separate quantum operations, $\mathcal{E}_A\left(\rho\right)$ and $\mathcal{E}_B\left(\rho\right)$ for the decoherence of Alice and Bob, were calculated using  $T_{\textrm{2E, A}}=10~\mu$s  and $T_{\textrm{2E, B}}=16~\mu$s respectively. The final density matrix, obtained by cascading the two operation, resulted in a $20\%$ infidelity due to decoherence, i.e $\mathscr{F}_{T_{2\textrm{Bell}}}\cong0.8$.

	\subsection{Dark Counts}

This protocol's robustness to loss is a result of heralding on single-photon detection events which are uniquely linked to the generation of a Bell state. However dark counts mix the Bell state with other states, $\ket{gg}$ for example, resulting in a lowered fidelity. This infidelity was calculated by modeling the impact of an imperfect detector on the two-qubit density matrix. The detector takes one of three possible input states, the flying Fock states $\ket{0}$, $\ket{1}$ and $\ket{2}$, and returns one of two outputs, click or no-click. In the generalized measurement formalism, this corresponds to the three measurement operators $\bm{M}_0=\ket{0}\bra{0}$, $\bm{M}_1=\ket{0}\bra{1}$ and $\bm{M}_2=\ket{0}\bra{2}$ for detecting $0$, $1$ or $2$ photons respectively\cite{Haroche2006}. To model the imperfections of dark counts and finite detector efficiency, we introduce $P_\textrm{d}$, the probability of a dark count in the detector, and $P_\textrm{real}$, the probability that the detector goes click when a photon arrives. Since according to simulations, the detector cannot distinguish between $\ket{1}$ and $\ket{2}$, we make the assumption that either input results in a click with the same probability $P_\textrm{real}$. Thus, the probability of the two outcomes, no-click (NC) and click (C), are:

\begin{align*}
P_\textrm{NC}=\textrm{Tr}\left[\left(1-P_\textrm{d}\right)\bm{M}_0\rho\bm{M}_0^\dagger+\left(1-P_\textrm{real}\right)\left(\bm{M}_1\rho\bm{M}_1^\dagger+\bm{M}_2\rho\bm{M}_2^\dagger\right)\right]
\end{align*}
\begin{align*}
P_\textrm{C}=\textrm{Tr}\left[P_\textrm{d}\bm{M}_0\rho\bm{M}_0^\dagger+P_\textrm{real}\left(\bm{M}_1\rho\bm{M}_1^\dagger+\bm{M}_2\rho\bm{M}_2^\dagger\right)\right] 
\end{align*}

Based on the measurement outcome, the input density matrix is projected to one of two output density matrices: 
\begin{equation}
\rho_{\textrm{NC}}=\frac{\left(\left(1-P_\textrm{d}\right)\bm{M}_0\rho\bm{M}_0^\dagger+\left(1-P_\textrm{real}\right)\left(\bm{M}_1\rho\bm{M}_1^\dagger+\bm{M}_2\rho\bm{M}_2^\dagger\right)\right)}{P_\textrm{NC}}
\end{equation}
\begin{equation}
\rho_{\textrm{C}}=\frac{\left(P_\textrm{d}\bm{M}_0\rho\bm{M}_0^\dagger+P_\textrm{real}\left(\bm{M}_1\rho\bm{M}_1^\dagger+\bm{M}_2\rho\bm{M}_2^\dagger\right)\right)}{P_\textrm{C}}
\end{equation}

To model the experiment and calculate the fidelity limited by dark counts, the final density matrix after two rounds of the protocol and successful photon detection was calculated, resulting in a $36\times36$ density matrix.  The photon components of the density matrix were traced out, yielding the $4\times4$ density matrix $\rho_\textrm{final}$. From this, the fidelity limited by dark counts, $\mathscr{F}_{\textrm{det}}=\textrm{Tr}\left(\rho_\textrm{final}\ket{O^+}\bra{O^+}\right)$, was found:

\begin{equation}
\mathscr{F}_{\textrm{det}}=\frac{3P_{\textrm{d,1}}P_{\textrm{d,2}}+P_{\textrm{d,1}}P_{\textrm{real,2}}+4P_{\textrm{real,1}}P_{\textrm{real,2}} }{11P_{\textrm{d,1}}P_{\textrm{d,2}}+8P_{\textrm{d,2}}P_{\textrm{real,1}}+9P_{\textrm{d,1}}P_{\textrm{real,2}}+4P_{\textrm{real,1}}P_{\textrm{real,2}}}
\end{equation}

Here the numeric subscripts on $P_\textrm{d}$ and $P_\textrm{real}$ are for the two detections rounds in the experiment. The values of $P_\textrm{real}$ and $P_\textrm{d}$ for each round were extracted from the measured click probabilities from the remote entanglement and the control experiments. We find $P_\textrm{d,1}=0.006, P_\textrm{d,2}=0.005, P_\textrm{real,1}=0.21, P_\textrm{real,2}=0.26$ and thus $\mathscr{F}_{\textrm{det}}\cong0.9$. Combining the effects of decoherence and dark counts results in an expected theoretical fidelity of $\mathscr{F}_{\textrm{thy}}=0.76$.

From this model of the experiment, it also possible to analyze the state created at the end of the first round of the protocol which, as described in the main text, is $\rho_3^{click}=\mathscr{N} \ket{O^+}\bra{O^+} + \left(1-\mathscr{N}\right) \ket{ee}\bra{ee}$. In the case of a detector with no ability to distinguish between the inputs $\ket{1}$ and $\ket{2}$ and with no losses or dark counts, the normalization constant $\mathscr{N}=\frac{2}{3}$. However, in the presence of dark counts in the detector, the mixed state is further contaminated with weights in $\ket{gg}$ and $\ket{O^-}$. From the values of $P_\textrm{d,1}$ and $P_\textrm{real,1}$, we find that the mixed state generated by detecting a click in the first round is:

\begin{equation}
\rho_3^{click}=0.635 \ket{O^+}\bra{O^+} + 0.327 \ket{ee}\bra{ee} + 0.019 \ket{gg}\bra{gg} + 0.019 \ket{O^-}\bra{O^-}
\end{equation}

While limiting the experiment to a single round would reduce the effects of decoherene and increase the generation rate by decreasing the protocol time, the advantages of performing a second round of the protocol are greater. In addition to increasing the overall fidelity by removing the weight in the $\ket{ee}$, the inclusion of the $R_y(\pi)$ pulse on Alice and Bob stabilizes the phase of the generated Bell state, protecting it against inevitable drifts in experimental setup, such as the qubit frequencies or the phase of the various microwave generators for example.

	\subsection{Tomography}

To model the imperfections arising from the tomography process, we used the theory described above (see Appendix B) to calculate the Pauli components in Fig.~4. Using the experimentally measured $\bm{A}$ matrix, the imperfect projectors $\Pi_j^{expt}$ were calculated. Thus the measurement outcome is $P_{jk}=\textrm{Tr}\left[\Pi_j^{expt}\bm{R}_k\rho\bm{R}_k^\dagger\right]$ where $\bm{R}_k$ is one the $9$ tomography pre-rotations. From the set of measurement outcomes, the $16$ Pauli vector were calculated and plotted in Fig.~4B.

	\subsection{Error Analysis}

The error bars in the quoted fidelity were dominated by statistical error from the finite number of tomography outcomes used to reconstruct the density matrix in the Pauli basis. Since around $2\times10^5$ successful shots were used to calculate each Pauli component, the error was limited to around $1\%$. To convert the error in the Pauli components to an error in the fidelity and hence the concurrence, different density matrices were constructed by varying each Pauli component by their respective error amounts. The fidelity to $\ket{O^+}$ and concurrence was then calculated for each of these density matrices to find the desired error bars.

\clearpage
\begin{acknowledgments}
The authors thank B. Vlastakis for helpful discussions and M. Rooks for fabrication assistance. Facilities use was supported by the Yale Institute for Nanoscience and Quantum Engineering (YINQE), the National Science Foundation (NSF) MRSEC DMR 1119826, and the Yale School of Engineering and Applied Sciences cleanroom. This research was supported by the U.S. Army Research Office (Grant No. W911NF-14-1-0011), and the Multidisciplinary University Research Initiative through the US Air Force Office of Scientific Research (Grant No. FP057123-C). W. P. was supported by NSF grant PHY1309996 and by a fellowship instituted with a Max Planck Research Award from the Alexander von Humboldt Foundation.
\end{acknowledgments}

%\bibliography{Fockbib}

%

\end{document}